\documentclass[%
 reprint,
superscriptaddress,
nofootinbib,
 amsmath,amssymb,
 aps,
]{revtex4-2}

\usepackage{graphicx}
\usepackage{dcolumn}
\usepackage{bm}
\usepackage{braket}

\usepackage[table]{xcolor}
\usepackage{float}
\usepackage{siunitx}
\usepackage{amsmath}
\usepackage{xcolor}
\usepackage{lipsum} 

\begin{document}

\preprint{APS/123-QED}


\title{Scalable Quantum Spin Networks from Unitary Construction}
 

\author{Abdulsalam H. Alsulami}%
\email{aha555@york.ac.uk}
 \affiliation{%
 York Centre for Quantum Technologies, Department of Physics, University of York, York, YO105DD
 }%
\author{Irene D'Amico}%
\email{irene.damico@york.ac.uk}
 \affiliation{%
 York Centre for Quantum Technologies, Department of Physics, University of York, York, YO105DD
 }%
\author{Marta P. Estarellas}%
\email{mpestarellas@qilimanjaro.tech}
 \affiliation{%
 Qilimanjaro Quantum Tech, Barcelona 08007, Spain
 }%
\author{Timothy P. Spiller}%
\email{timothy.spiller@york.ac.uk}
 \affiliation{%
 York Centre for Quantum Technologies, Department of Physics, University of York, York, YO105DD
 }%

\date{\today}

\begin{abstract}
Spin network systems can be used to achieve quantum state transfer with high fidelity and to generate entanglement. A new approach to design spin-chain-based spin network systems, for short-range quantum information processing and phase-sensing, has been proposed recently in \cite{alsulami2022unitary}. In this paper, we investigate the scalability of such systems, by designing larger spin network systems that can be used for longer-range quantum information tasks, such as connecting together quantum processors. Furthermore, we present more complex spin network designs, which can produce different types of entangled states. Simulations of disorder effects show that even such larger spin network systems are robust against realistic levels of disorder.
\end{abstract}

\maketitle

\section{\label{sec:intro}Introduction}
Spin chain systems are linear Spin Networks (SN) that have been shown to be useful for realising various tasks in quantum information processing (QIP). For example, such systems can be used to efficiently transfer quantum information from site to site, a process known as quantum state transfer \cite{bose2007quantum,nikolopoulos2004electron,serra2021pretty}. Transferring information with unit fidelity can occur in certain SN, in a process known as Perfect State Transfer (PST) \cite{christandl2004perfect,burgarth2005conclusive,vinet2012construct}. SN have interesting entanglement properties \cite{ramirez2015entanglement} and can also be used to generate and/or distribute entanglement \cite{d2007freezing,estarellas2017robust}.

One of the main advantages of SN systems is that they represent a generic mathematical model of coupled two-level systems, that can be realised experimentally by means of different physical systems, such as quantum dots \cite{loss1998quantum,nikolopoulos2004electron,d2006quantum}, trapped ions \cite{brown2016co}, and superconducting qubits \cite{berggren2004quantum,karamlou2022quantum}. Another advantage is that SN can be set up to achieve desirable QIP tasks through natural dynamics \cite{mortimer2021evolutionary}, without requirement of switching on or off couplings between qubits (e.g. using time-dependent fields), which otherwise may cause additional errors.

Spin systems with more complex topologies \cite{ChristandlMatthias2005Ptoa,KayAlastair2011Bopc,pemberton2011perfect,roy2018response,riegelmeyer2021generation,alsulami2022unitary} have potential for wider application than linear chain systems, including for quantum sensing \cite{degen2017quantum,pirandola2018advances,alsulami2022unitary}. A simple example of a SN system, designed by coupling together two identical PST chains via a unitary transformation, has already been proposed, \cite{alsulami2022unitary}. These systems were shown to have applications for robust routing of quantum information, entanglement generation and quantum phase sensing. Here, we systematically apply the unitary construction method in Ref \cite{alsulami2022unitary} to show its potential to produce scalable SN systems. This larger SN systems could be used for the application of longer-range QIP tasks. Furthermore, more complex forms of SN are proposed in order to realise different types of bipartite and multipartite entanglement. Modelling of different types of disorder effects in these systems reveals that they function very robustly under such errors, demonstrating the potential to be used for various practical quantum technology purposes.

The outline of this paper is as follows: section \ref{model sec} will describe the model of our system. Section \ref{2-chains SN} will discuss scalable SN systems, designed by coupling together two PST chains, that can be used for various QIP purposes. Section \ref{Multi-chains SN} will introduce more complex SN systems, designed by coupling together multiple PST chains, with its applications. Then, in Section \ref{Conclusion}, we conclude and discuss future work. 

\section{The Model}
\label{model sec}
A generic SN system can be described by the following time-independent XY-Hamiltonian
\begin{equation}      
\label{Eq:SC Hamiltonian}
\mathcal{H}_{XY} = \frac{1}{2} \sum_{i,j} J_{i,j} (\sigma_i^x \sigma_{j}^x + \sigma_i^y \sigma_{j}^y ) 
+ \sum_{i=1}^{N} \frac{\epsilon_i}{2} (\sigma_i^z) 
\end{equation} 
where $N$ is the total number of sites/spins and $J_{i,j}$ are the coupling 
parameters representing the interaction between sites $i$ and $j$. The spin component for site $i$ is represented using Pauli operators $ \sigma^x_i $, $ \sigma^y_i $, $ \sigma^z_i $. The second term in the equation represents the on-site energies for all sites $i$ (the energy cost for an excitation at site $i$). Unless diagonal error is included in the system, we consider the on-site energies to be independent of the site $i$, so we can set all these to zero,  $\epsilon_i=0$. 

In a SN prepared such that all sites have spin down, a single-excitation at a site $i$ is defined as a spin up, with a state $\ket{r_i}=\ket{00\dots1_i00\dots}$. The total number of excitations is conserved, as the Hamiltonian $\mathcal{H}_{XY}$ commutes with the total number of excitations, even in the presence of disorder. We will therefore only consider the single-excitation sub-space, as this suffices to achieve the desired QIP phenomena.   

It is well-known that spin chains can exhibit PST or quasi-PST by setting the coupling parameters $J_{i,i+1}$ either by tuning the boundary couplings, $J_{1,2}$  and $J_{N-1,N}$ \cite{wojcik2005unmodulated,oh2012effect,banchi2010optimal,banchi2011nonperturbative}, or by controlling each coupling parameter, $J_{i,i+1}$ \cite{kay2010perfect,karbach2005spin,kostak2007perfect, christandl2004perfect,nikolopoulos2004electron,ChristandlMatthias2005Ptoa}. We use the latter approach, where the coupling parameters are set to be
\begin{equation}
\label{Eq.controlled coupling}     
    J_{i,i+1}=J_0 \sqrt{i(N-i)} \;.
\end{equation}    

with the maximum coupling $J_{max}$ occurring in the middle of the chain. We set $J_{max}=1$ (unless otherwise stated), to define our energy units. Since the maximum coupling occurs in the middle of the chain, we can derive $J_0$ as $J_0 = 2J_{max}/N$ and $J_0 = J_{max}/\sqrt{\frac{N^2}{4}-\frac{1}{4}}$, for even and odd chains respectively. 

Fidelity is a useful tool used to test how well a desirable process is achieved. Here it will track the overlap in time of the system state with a desirable state $\ket{\psi_{des}}$
\begin{equation}
\label{equationFidelity Eq}
F(t) = |\bra{\psi_{des}}e^{-i\mathcal{H}t}\ket{\psi(0)}|^2,
\end{equation}
with $\ket{\psi(0)}$ the initial state, the relevant static system Hamiltonian $\mathcal{H}$ and with the reduced Plank constant set to be $\hbar=1$. 

For a chain described by Eq. (\ref{Eq.controlled coupling}), PST happens at time $t_m=\pi/(2J_0)$ (also known as mirroring time)  \cite{ChristandlMatthias2005Ptoa}, and is achieved when the fidelity of an evolved initial state $\ket{r_1}=\ket{100\ldots}$ against the desirable state $\ket{r_N}=\ket{00\ldots01}$ is equal to unity. Note that, given the relationship between $J_{0}$ and $J_{max}$, for a fixed $J_{max}=1$ the mirror time scales in proportion to $N$, which is the natural situation for any physical implementation.

As will be discussed shortly, to describe practical systems, we consider errors. We therefore use ensembles of systems described by Hamiltonians containing independent random errors. For such ensembles, the average fidelity can be calculated as 
\begin{equation}
    \overline{F}(t) = Tr(\rho(t)\ket{\psi_{des}}\bra{\psi_{des}}),
\end{equation}
where  $\rho(t)=\frac{1}{K}\sum_{i=1}^{K}\ket{\psi_i(t)}\bra{\psi_i(t)}$ is the ensemble density matrix and $K$ the number of systems in the ensemble.

Entanglement of Formation ($EOF$) is used to quantify the degree of entanglement between a pair of qubits regardless of whether they are in a pure or mixed (reduced) state. The $EOF$ is defined as in Ref. \cite{wootters2001entanglement}. When considering the presence of random errors in the system, the average of many realisations of $EOF$ is calculated as $\overline{EOF}=\frac{1}{K}\sum_{i=1}^{K}f(\rho^{red})_i$, where $f(\rho^{red})_i=EOF_i$ represents the $EOF$ calculated from the reduced density operator $\rho^{red}$ of the \textbf{two} relevant sites for a single randomly generated example of disorder. \footnote{We note that this differs from \cite{alsulami2022unitary}, where the average of many realisations of the reduced density operator is calculated first, $\overline{\rho^{red}}=\frac{1}{K}\sum_{i=1}^K\rho^{red}_i$, and the average $EOF$ is then calculated as $EOF=f(\overline{\rho^{red}})$.} 


\subsection{Off-Diagonal Disorder}
Off-diagonal disorder represents the error in the coupling parameters of the system. The effect of this type of error on the system is therefore investigated by adding a perturbation to the coupling interaction of the Hamiltonian $\mathcal{H}_{XY}$ in Eq.~(\ref{Eq:SC Hamiltonian}) 
\begin{equation} 
\label{equationoff-diag error} 
J_{i,j}^{perturbed} = J_{i,j} + J_{i,j}^{'} \;,
\end{equation}
where $J_{i,j}^{'} = E d_{i,j}J_{max}$. The perturbation units are set by $J_{max}$, the dimensionless parameter $E$ sets the scale of the error, and $d_{i,j}$ represent random numbers, here from a Gaussian distribution with a standard deviation of $w=\frac{1}{2\sqrt{3}}$ and a zero mean, \cite{alsulami2022unitary}. 

\subsection{Diagonal Disorder}
The diagonal disorder represents the inhomogeneity in the energy required to excite a site $i$. Therefore, the second term of the Eq.~(\ref{Eq:SC Hamiltonian}) is now present, with $\epsilon_i = Ed_{i}J_{max}$, where $E$ and the random $d_{i}$ have the same definitions as before.

\section{Two-chain Spin Networks}
\label{2-chains SN}
Various quantum processing operations, such as routing, entanglement, and phase sensing have been implemented previously in a small SN of size $N=6$ \cite{alsulami2022unitary}, and in order to account for scalability we will now investigate these phenomena in larger SN systems. 

\begin{figure}[ht!]
\centering
\includegraphics[width=0.45\textwidth]{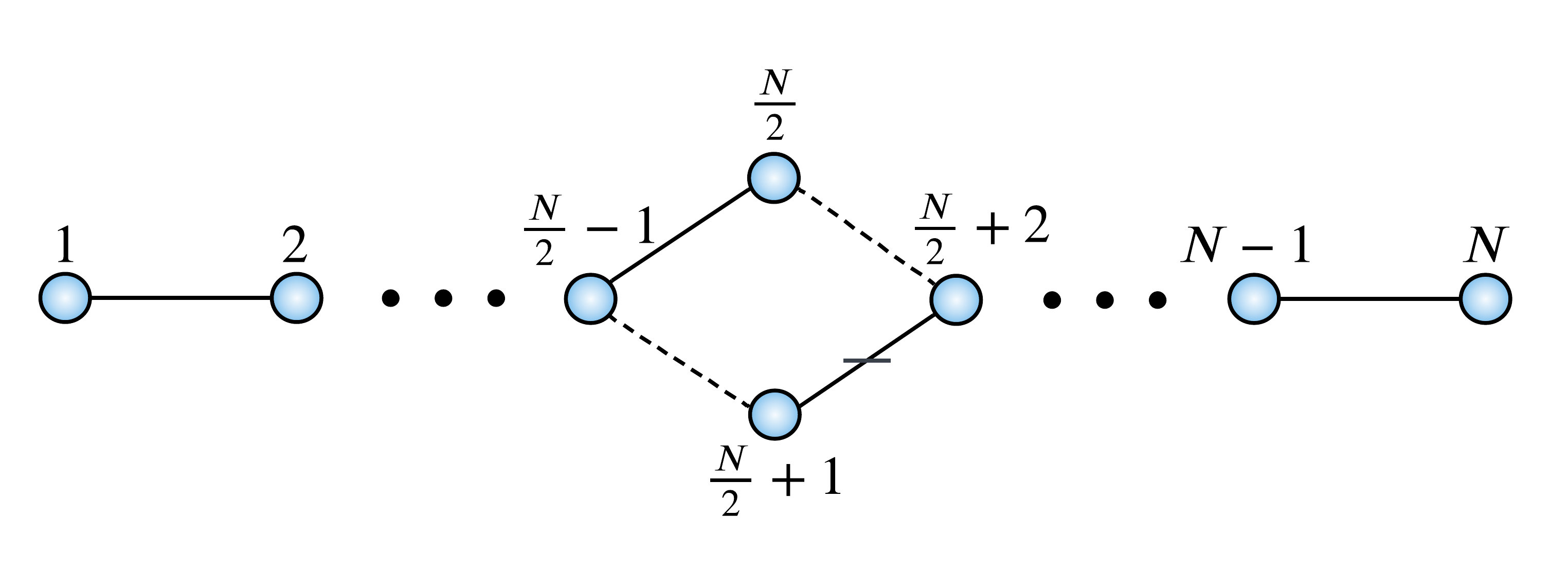}
\caption{Diagram of a large two-chain SN system of size $N$.}
\label{SN device}     
\end{figure}

One way to scale the SN system is by connecting together longer spin chains. 
Following the method in \cite{alsulami2022unitary}, we design an, in principle, arbitrarily large SN system, by coupling together two PST chains by means of a Hadamard-like unitary transformation, \cite{alsulami2022unitary}. Such a SN will maintain the PST property within each original chain, as the unitary transformation of the uncoupled chains preserves the spectrum and only changes the eigenstates. An example of a large two-chain SN, each of size $N/2$ is shown in Figure.\eqref{SN device}. We stress that the unitary transformation is a design step, and so it is the final SN device, Figure.\eqref{SN device}, that should be implemented experimentally. Note, the dashed lines in Figure.\eqref{SN device} are the new couplings that connect the two uncoupled chains as a result of the unitary transformation of the Hamiltonian. Furthermore, the coupling between sites $\frac{N}{2}+1$ and $\frac{N}{2}+2$ is now negative as denoted by the horizontal bar. This notation for couplings applies to all SN figures. 

This and other spin-chain-based SN can be used for longer-distance quantum communication, or longer-range entanglement generation than the spin-chain-based SN in \cite{alsulami2022unitary}. Such SN systems have an advantage over spin chains due to the fact that they offer wider opportunities for routing, and can be used at the same time also for phase sensing and entanglement generation \cite{alsulami2022unitary}. Moreover, for a SN of multiple chains, discussed in Section \ref{Multi-chains SN}, the difference in energy between the largest and the smallest coupling, for an equivalent long spin chain, would be larger, and this could be an experimental limitation, depending on hardware. Similarly, connecting together a few short chains instead of utilising just a single longer one, may reduce the number of different values for the couplings to be experimentally engineered, which may be an advantage for certain types of implementation. 

\subsection{Router between qubits $1$ and $N$}
One of the major advantages of using a spin-chain-based SN is that useful phenomena can be realised by exploiting the natural dynamics of the system. Considering the SN system of size $N$ (even), Fig.\eqref{SN device}, the system can be operated as a router such that information (encoded as a single-excitation) is sent from site 1 to site $N$, as we now describe.

\begin{figure}[ht!]
    \centering
    \includegraphics[width = 0.45\textwidth]{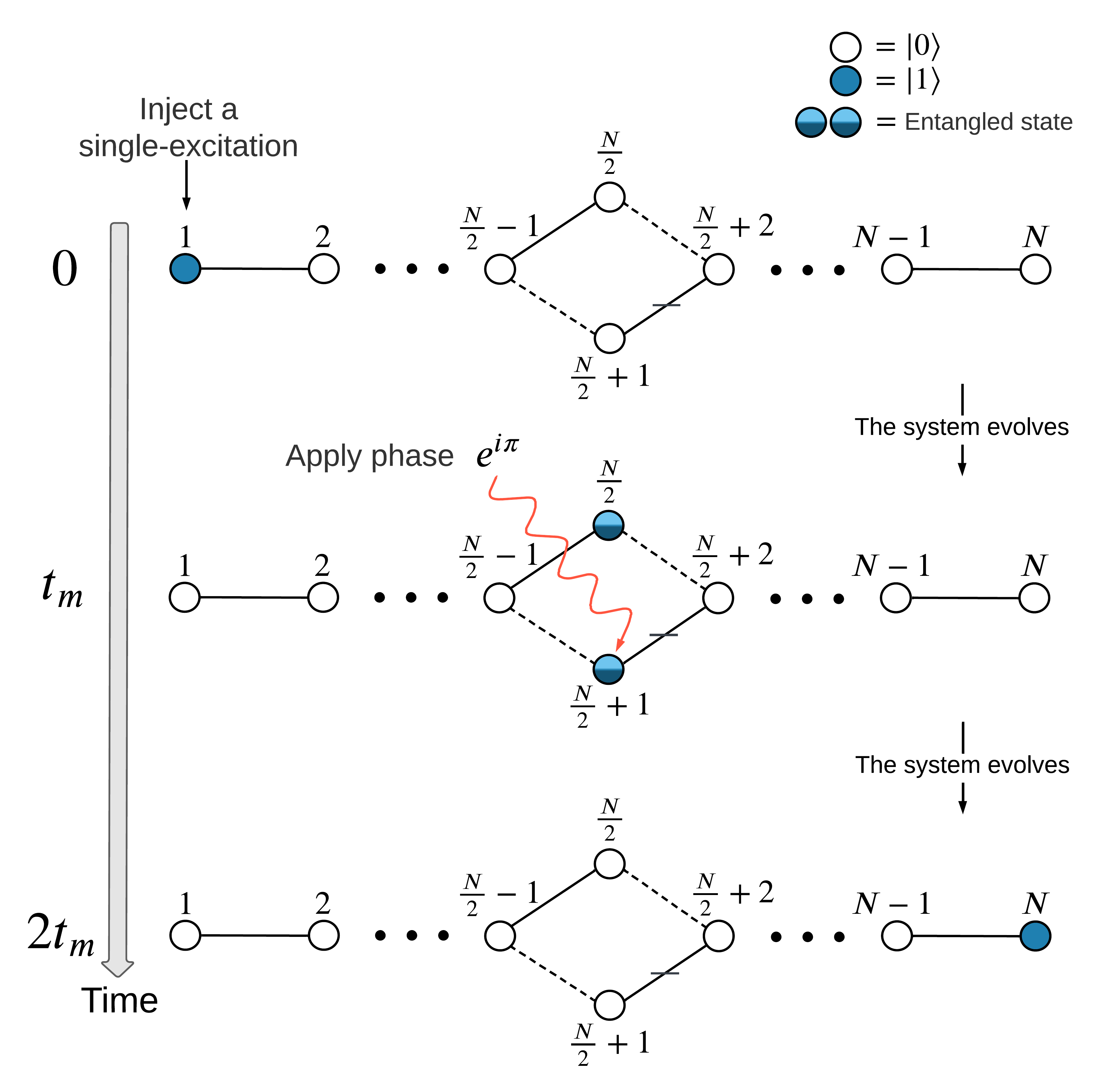}
    \caption{The routing protocol is achieved by injecting a single-excitation at site 1 at $t=0$ and a phase flip at site $\frac{N}{2}+1$ at $t=t_m$.}
    \label{router protocol}     
\end{figure}

\subsubsection{Router Protocol}
We start by first preparing the system such that all sites have spin down state $\ket{00\ldots 0}$. Then, a single-excitation is injected at site 1 at $t=0$ 
\begin{equation}
\label{equationexcit at site1 SN}   
\ket{\psi(0)} = \ket{r_1}, 
\end{equation}

Under the natural dynamics of the system, this state will evolve through the system as $\ket{\psi(t)}=e^{-i\mathcal{H}t}\ket{\psi(0)}$. At the mirroring time, $t=t_m$, the excitation will be in a superposition between sites $\frac{N}{2}$ and $\frac{N}{2}+1$

\begin{equation}
\label{equations3&s6} 
\ket{\psi(t_m)} = \frac{e^{-i\varphi(N)}}{\sqrt{2}}(\ket{r_{\frac{N}{2}}} + \ket{r_{\frac{N}{2}+1}}).
\end{equation} 

The overall phase factor is given by $e^{-i\varphi(N)}=(-i)^{\frac{N}{2}-1}$. This is consistent with previous results for linear chains \cite{ChristandlMatthias2005Ptoa}. The state of the system then evolves back to site 1 at $t=2t_m$. This is because of the SN PST properties and the Hadamard-based construction. 

In order to operate the system as a router, we intervene by applying a phase flip at either site $\frac{N}{2}$ or site $\frac{N}{2}+1$ at $t=t_m$. We choose to apply the phase flip at site $\frac{N}{2}+1$, obtaining

\begin{equation}
\label{equations3&es6} 
\ket{\psi(t_m)}_{\pi} = \frac{e^{-i\varphi(N)}}{\sqrt{2}}(\ket{r_{\frac{N}{2}}} + e^{i\pi}\ket{r_{\frac{N}{2}+1}})\;.
\end{equation}
The state will then evolve to site $N$ at $t=2t_m$, so
\begin{equation}
\label{router state}   
\ket{\psi(2t_m)} = e^{-i\gamma(N)}\ket{r_N}\; , 
\end{equation}
with an overall phase factor of $e^{-i\gamma(N)}=(-i)^{N-2}$.

The routing protocol is described in Fig.\eqref{router protocol} and is confirmed by our numerical simulations.

\subsubsection{Router Robustness} 
\label{2_equal_ch_SN_router_robustness}
\begin{figure}[ht!] 
    \centering
    \includegraphics[width = 0.45\textwidth]{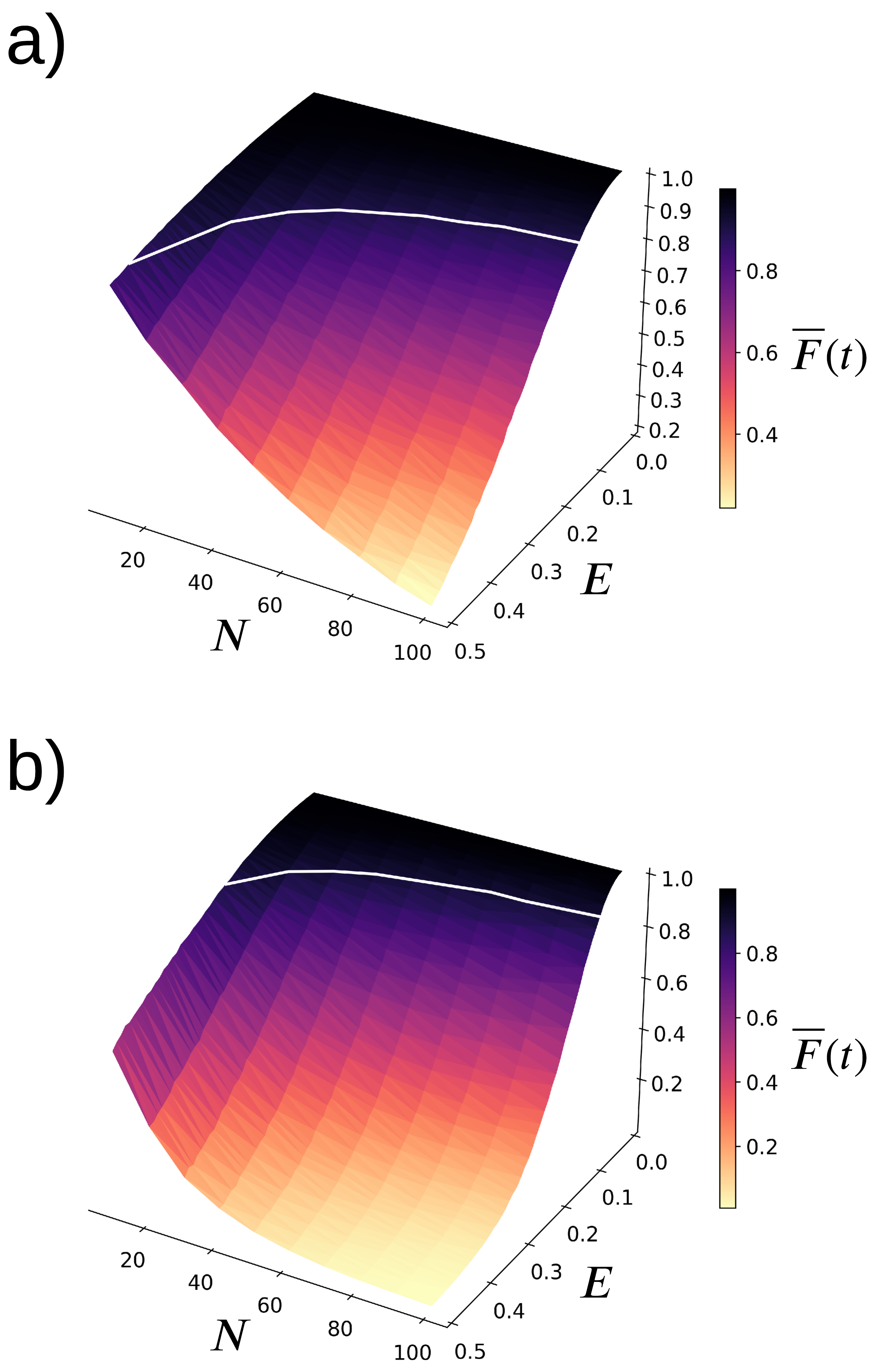} 
    \caption{The router fidelity robustness at $t=2t_m$ in the presence of diagonal (a) and off-diagonal (b) disorder against different $N$ values and for different error strength $E$. The white line indicates the fidelity $90 \%$ threshold ($\pm 1\%$ due to numerical discretization).}
    \label{router robustness vs N and vs disorders}
\end{figure}

We now investigate the robustness of the routing protocol for different $N$ values. In order to do so, we measure the fidelity of the system against state $\ket{r_N}$ at $t=2t_m$, the expected time the excitation should take to evolve from site 1 to site $N$ in the error-free case. This is done for different error strength $E$ and for different $N$ values, as shown in Fig.\eqref{router robustness vs N and vs disorders}. Each point in the plots has been averaged over 1000 realisations. 

Interestingly, the routing fidelity at $t=2t_m$ in the presence of diagonal disorder remains above $98\%$ with error strength of $E\leq5\%$ up to $N=100$. Even for error strength up to $E=10\%$, which is relatively high error, the fidelity remains above $92\%$ up to $N=100$. On the other hand, the fidelity in the presence of off-diagonal disorder remains above $90\%$ for error strength of $E\leq10\%$ and up to $N=40$. The fidelity then decays as $E$ and $N$ increases. The system robustness for the routing protocol for large SN systems suggests that this SN design could be useful for quantum communications within the scale of a quantum processing device.

As we have seen above, the quality of the routing protocol degrades with an increasing amount of error in the system and with an increasing system size ($E$ and $N$, respectively). This is because the quantum information tasks the SN are designed to deliver rely on quantum interference between different amplitudes in the system, and so disruption of the interference damages delivery of the task. For most processes considered in this work, the value of the  performance metric (for example of the fidelity) in the ideal case is at its maximum: this implies that small modifications of the system parameters will affect the performance metric only to second order. It follows that increasing the number of couplings, keeping the average amount of disorder small, will thus affect the system performance less than increasing the average amount of disorder beyond a perturbative amount, while keeping $N$ constant. Further increasing disorder will additionally disrupt the coherence and quantum interference in the system, which are responsible for delivering the desired SN operation. Eventually, a very large amount of disorder can even result in localisation, where in effect the excitation does not move at all \cite{ronke2016anderson}. Furthermore, for large $N$, the system is more susceptible to error because there are more amplitudes involved in the quantum interference that deliver the process. This explanation of how the quality of the protocol degrades with disorder and with the scalability (increasing $N$) applies to all the protocols discussed in this paper.

\subsection{Entanglement Generation}
Quantum entanglement has been generated by different protocols based on spontaneous emission \cite{moehring2007entanglement,stephenson2020high} or coincident detection of two fiber-based infrared photons \cite{krutyanskiy2023entanglement}. Another approach with photons is the generation of mode entanglement, for example between the two output modes of a beam-splitter when just a single photon is sent into one of the input modes. In our work here we generate this form of entanglement. Such a bipartite maximally entangled state can  be generated between the ends of the SN in two different ways, as demonstrated below. 

\begin{figure}[ht!]
    \centering
    \includegraphics[width = 0.45\textwidth]{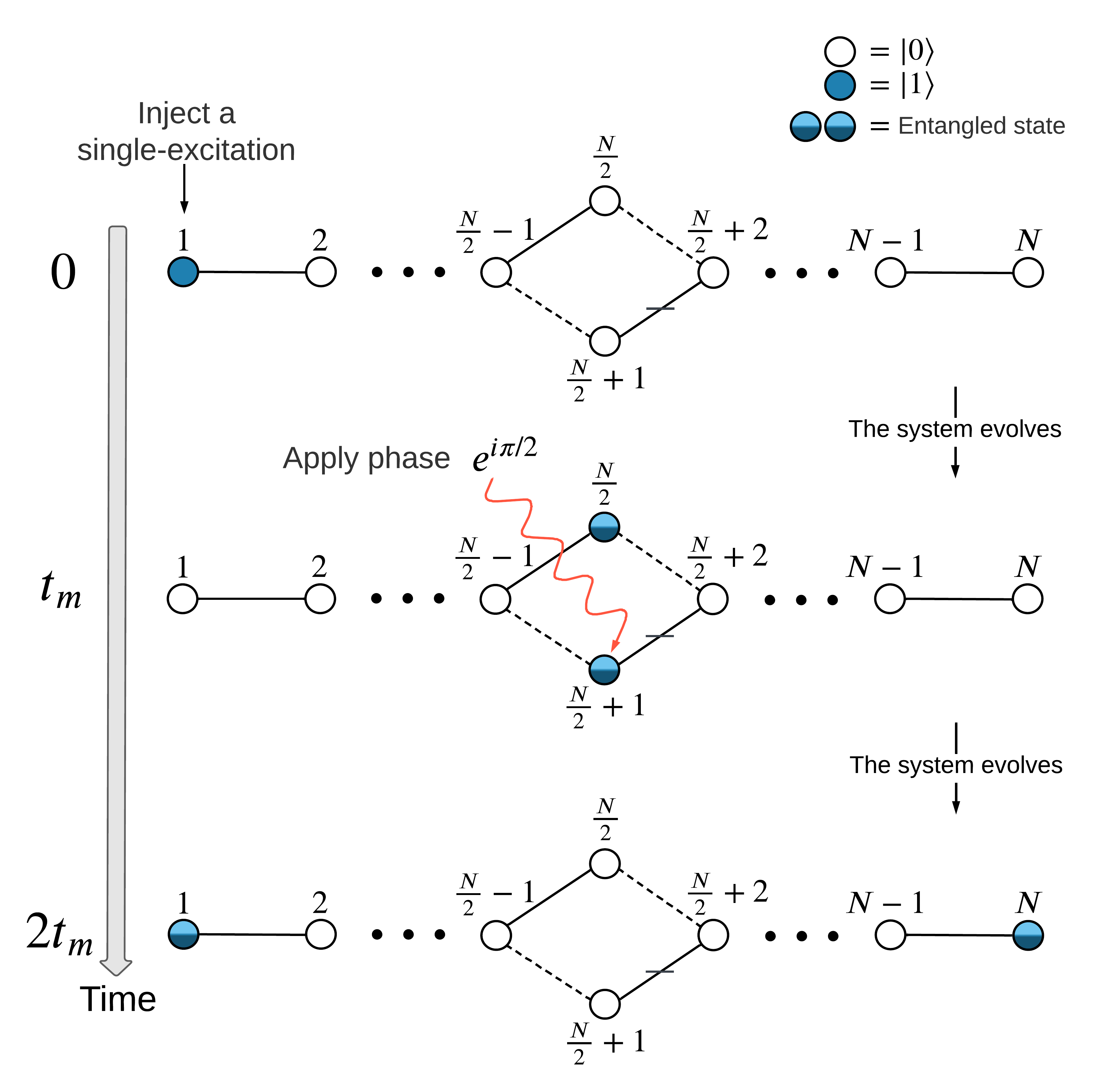}
    \caption{The entanglement generation protocol is achieved by injecting a single-excitation at site 1 at $t=0$ and a phase of $e^{i\pi/2}$ at site $\frac{N}{2}+1$ at $t=t_m$.}     
    \label{Entanglement protocol}     
\end{figure}

\subsubsection{First Entanglement Protocol}
The system is first initialised with a single-excitation injected at site 1 at $t=0$, as shown in Eq.\eqref{equationexcit at site1 SN}, and left to evolve. At $t=t_m$ the system will be in a state as given in Eq.\eqref{equations3&s6}, and we apply a phase of $e^{i\pi/2}$ at site $\frac{N}{2}+1$. The system is then left to evolve for another duration of $t_m$, which will result in a bipartite maximally entangled state at $2t_m$ given by
\begin{equation}
\label{Entang Eq}
\ket{\psi(2t_m)} = e^{-i\delta(N)}(\frac{1+e^{i\pi/2}}{2}\ket{r_1}+\frac{1-e^{i\pi/2}}{2}\ket{r_N}), 
\end{equation} \;
with a global phase given as $e^{-i\delta(N)}=(-1)^{\frac{N}{2}-1}$.
This entanglement generation protocol is shown in Fig.\eqref{Entanglement protocol}.

\begin{figure}[ht!] 
    \includegraphics[width = 0.45\textwidth]{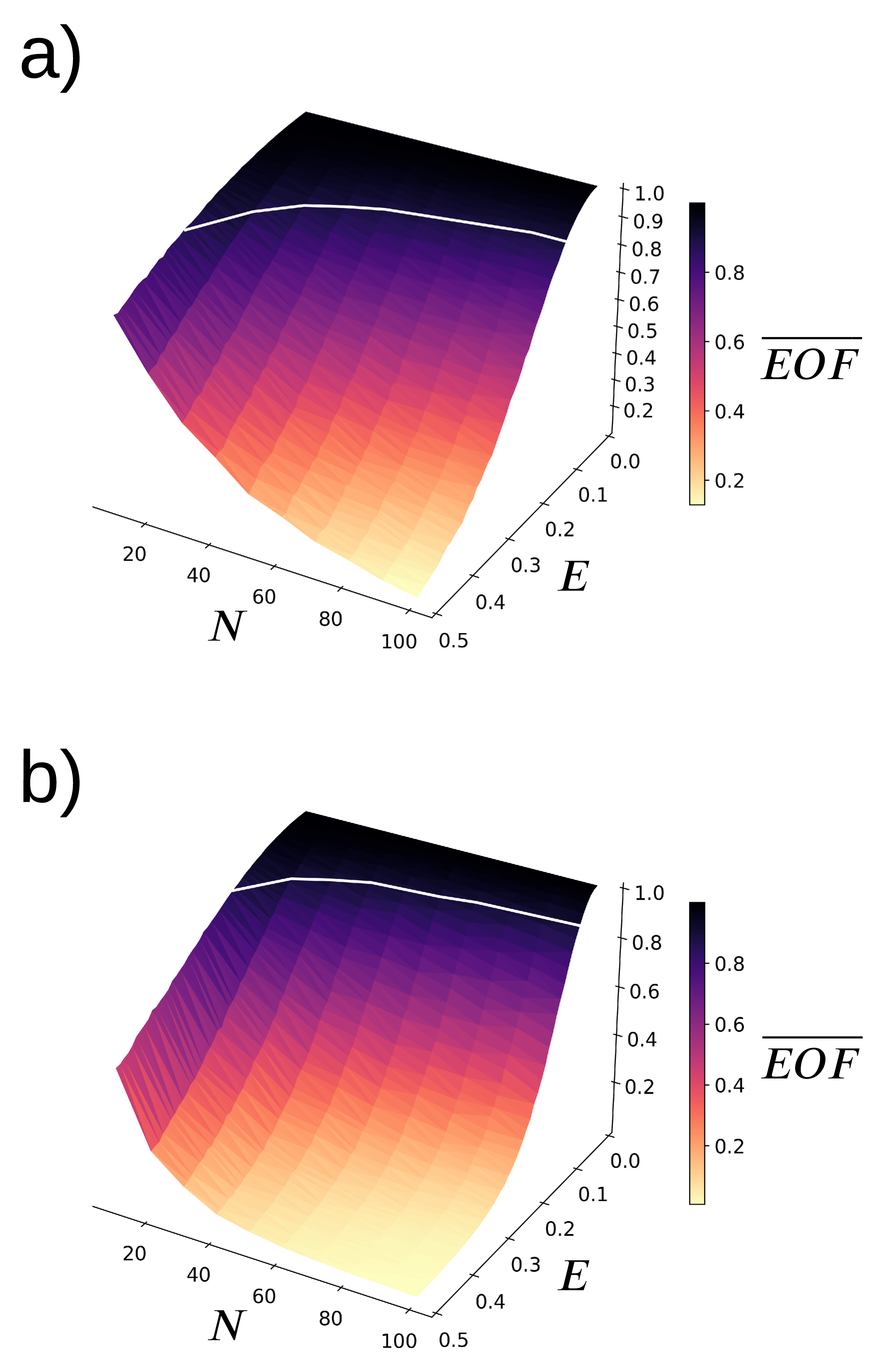} 
    \caption{$\overline{EOF}$ between sites 1 and $N$ at $t=2t_m$ for different SN size in the presence of diagonal (a) and off-diagonal (b) disorder with different error strength $E$. White lines have same meaning as in Fig.~\ref{router robustness vs N and vs disorders} but for the $\overline{EOF}$.}       
    \label{EOF robustness vs N and vs disorders}     
\end{figure}

We now investigate the robustness of this protocol. We therefore compute the $\overline{EOF}$ between sites 1 and $N$ at $t=2t_m$, the expected time at which the bipartite maximally entangled state forms in the error-free case. This is done for different error strengths $E$ and for different $N$ values, as shown in Fig.\eqref{EOF robustness vs N and vs disorders}. Each point in the plots has been averaged over 1000 realizations.

The $\overline{EOF}$ of the bipartite maximally entangled state in the presence of diagonal disorder remains above $96\%$ with $E=5\%$ and up to $N=100$. For a larger error of $E=10\%$ the $\overline{EOF}$ remains above $94\%$ up to $N=50$. On the other hand, the $\overline{EOF}$ is less robust  in the presence of off-diagonal disorder, with the $\overline{EOF}$  $\stackrel{>}{\sim}92\%$ for $E=10\%$ and SN size up to $N=20$. As the SN size $N$ and the error strength $E$ increases, the $\overline{EOF}$ decays.  

\subsubsection{Second Entanglement Protocol}
\label{seconed entang protocol}
In this protocol we initialise the system by injecting a single excitation at a central vertex of the diamond (e.g., at site $N/2$) such that
\begin{equation}
    \ket{\psi(0)}=\ket{r_{\frac{N}{2}}}. 
\end{equation}

Then the resultant evolution of this state at $t=t_m$ will be given by
\begin{equation}
    \ket{\psi(t_m)}=\frac{e^{-i\varphi(N)}}{\sqrt{2}}(\ket{r_1}
    +\ket{r_N}).
\end{equation}

This is a bipartite maximally entangled state between site 1 and site $N$. It  is achieved naturally, with no phase application. If not extracted or used, the state evolves back to the initial state, with the excitation localised at site $N/2$, at $t=2t_m$, to keep oscillating with a $2t_m$ period. A similar behaviour can be seen by injecting the single-excitation at site $\frac{N}{2}+1$, but with a different global phase due to the negative sign of the coupling between sites $\frac{N}{2}+1$ and $\frac{N}{2}+2$ \cite{alsulami2022unitary}.

\begin{figure}[H] 
    \includegraphics[width=0.45\textwidth]{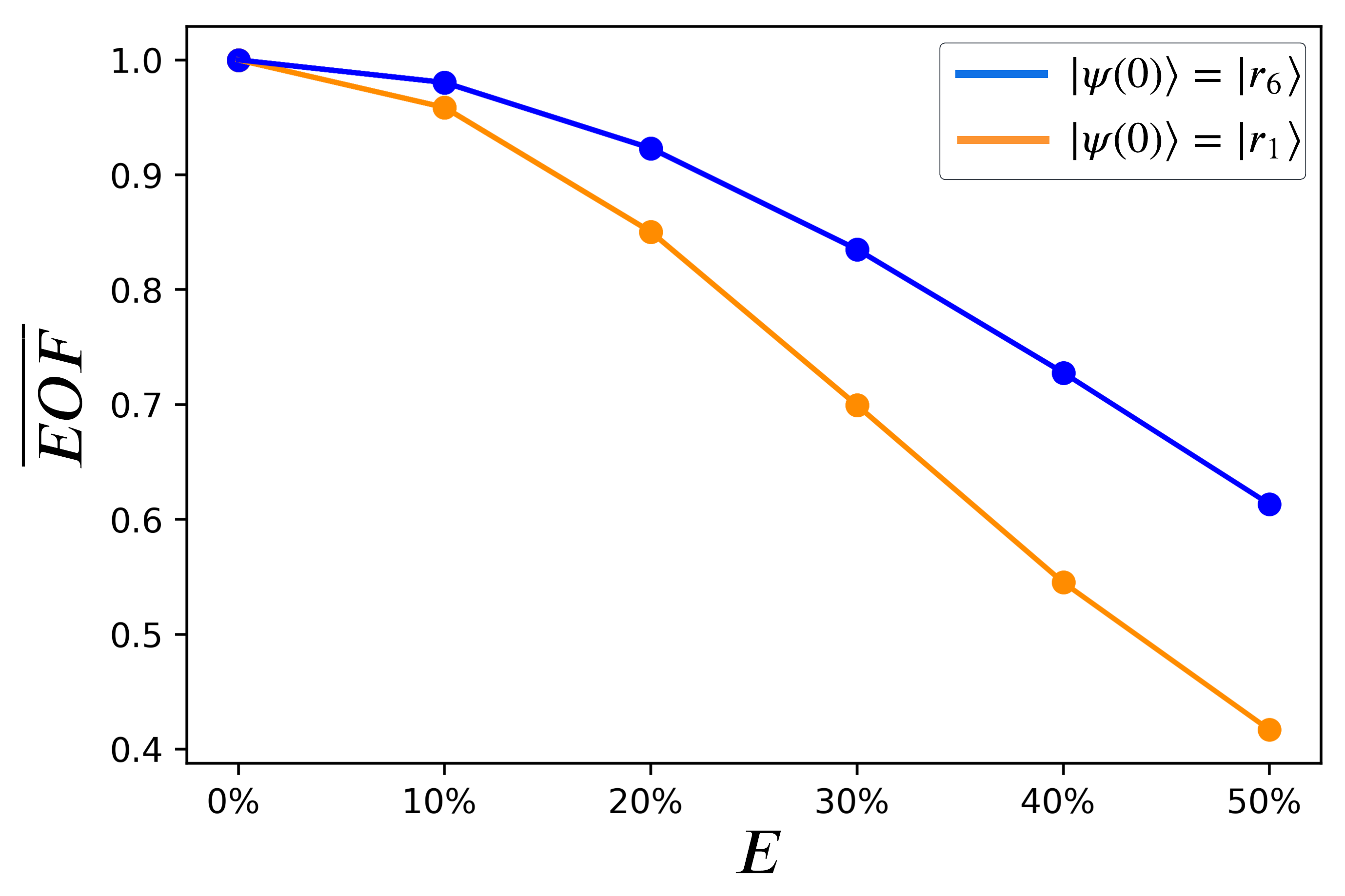} 
    \caption{Robustness of the bipartite maximally entangled state between site 1 and site $N$ (here $N=12$) for two different entanglement protocols in the presence of off-diagonal disorder. Blue: protocol with initial state $\ket{\psi(0)}=\ket{r_{6}}$, with $\overline{EOF}$ measured at $t=t_m$. Orange:  protocol with initial state $\ket{\psi(0)}=\ket{r_{1}}$, with $\overline{EOF}$ measured at $t=2t_m$.}
    \label{entanglement protocols compar}   
\end{figure}

We now compare the robustness of this  second entanglement protocol with the previous entanglement protocol. 
Fig.\eqref{entanglement protocols compar} describes a 12-site SN and shows that the second entanglement protocol, where $\ket{\psi(0)}=\ket{r_{\frac{N}{2}}}$, is more robust than the previous entanglement protocol (where $\ket{\psi(0)}=\ket{r_{1}}$), and especially so for $E>10\%$.  This is because the state evolution in the second entanglement protocol involves fewer sites, see Fig.\eqref{Colormap, n=6}, and is collected at $t=t_m$. Whereas the previous entanglement protocol involves all the sites with respect to the excitation evolution and is collected at $t=2t_m$. 
\begin{figure}[H] 
    \includegraphics[width=0.45\textwidth]{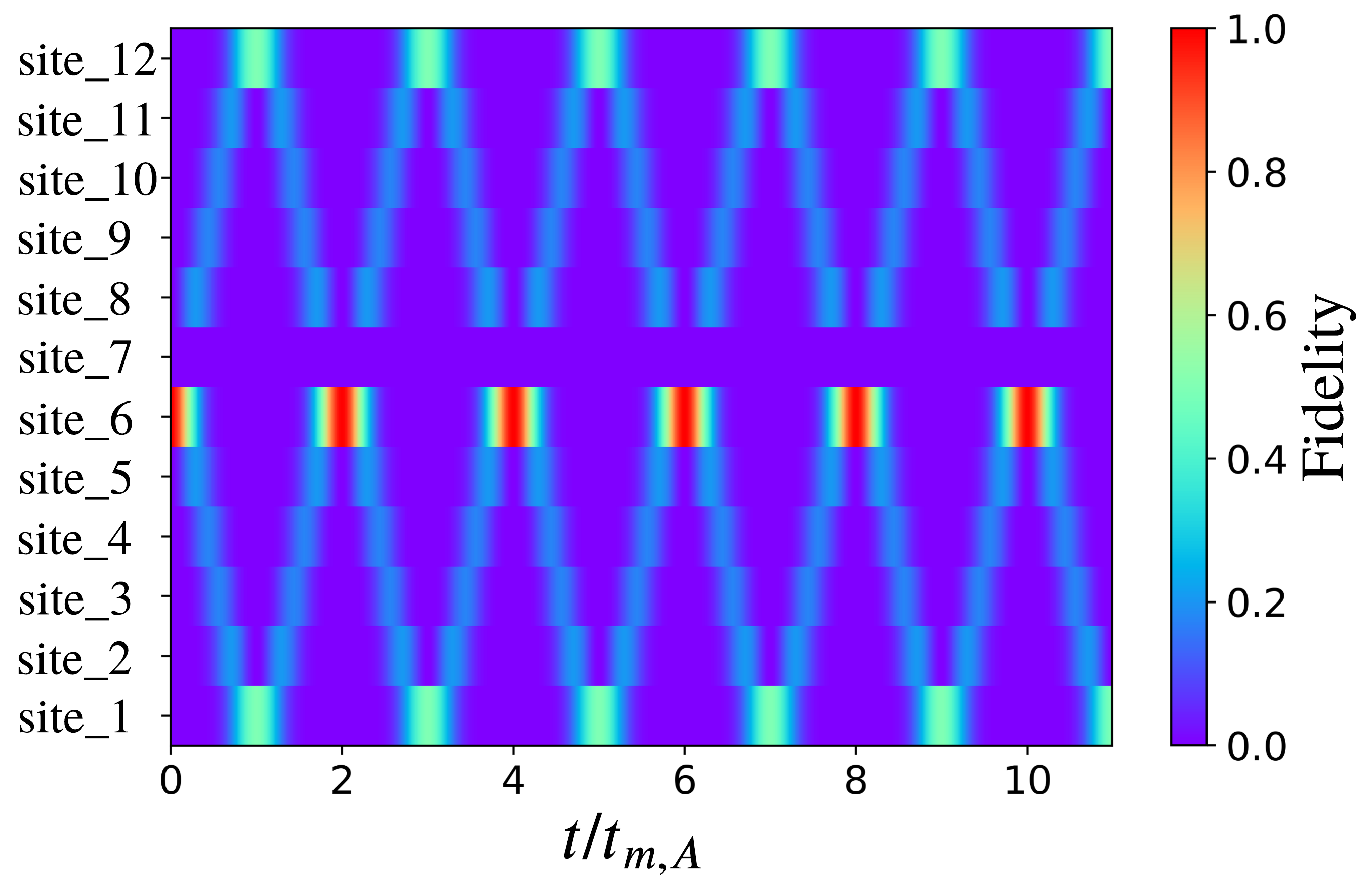} 
    \caption{Fidelity of each site as a function of the rescaled time $t/t_m$ in the second entanglement protocol. It shows that the excitation evolves through all sites except site 7. This is for SN of $N=12$.} 
    \label{Colormap, n=6}    
\end{figure}

\begin{figure*}[ht!]
    \centering
    \includegraphics[width=0.8\textwidth]{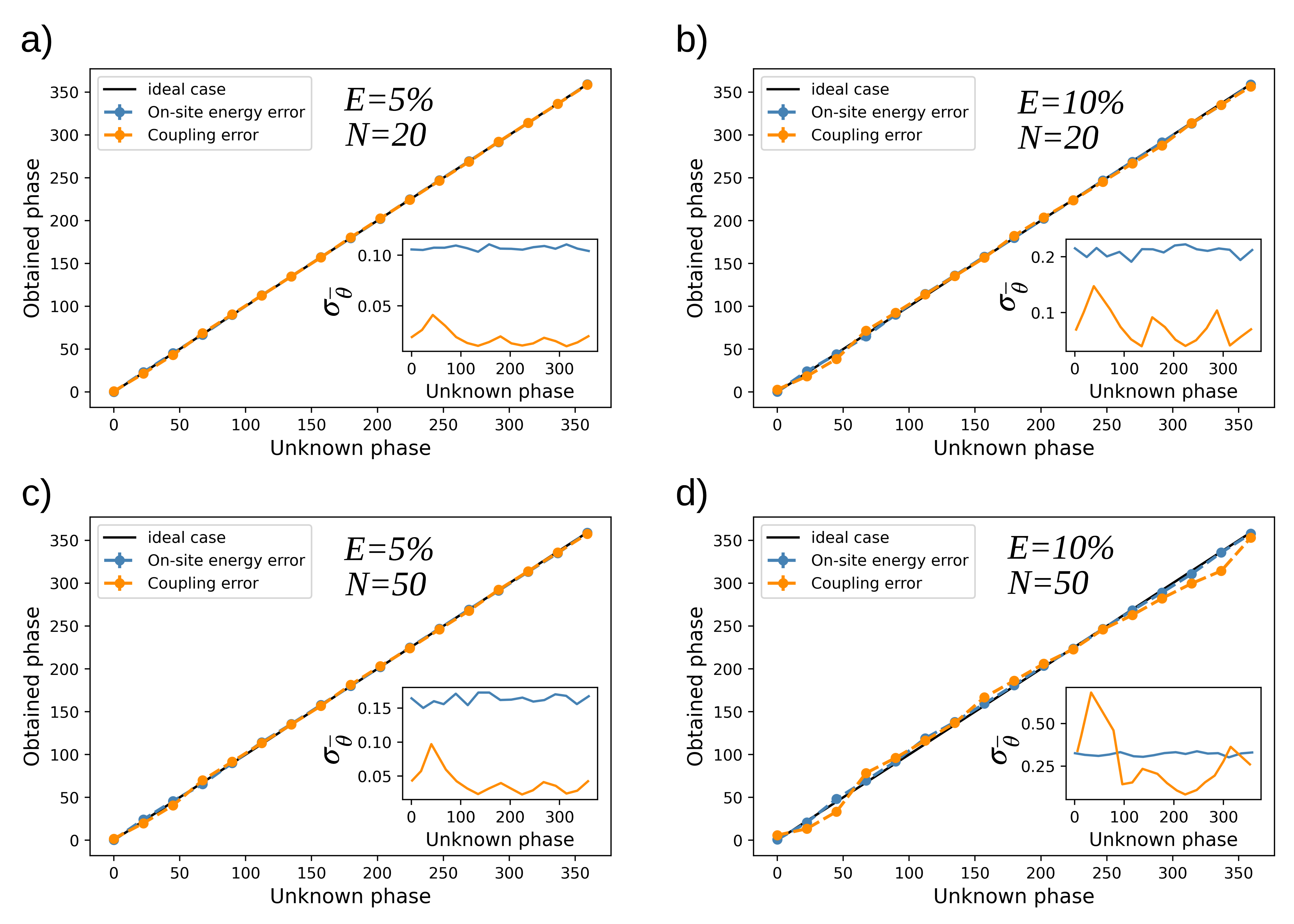}
    \caption{The obtained angles vs the unknown angles, in degrees, for SN of size $N=20$ with error strength of $E=5\%$ (a), for SN of size $N=20$ with error strength of $E=10\%$ (b), for SN of size $N=50$ with error strength of $E=5\%$ (c), and for SN of size $N=50$ with error strength of $E=10\%$ (d). This is shown also for the ideal case (no error, black) and for both types of disorder with angles being averaged over 1000 realisations. Inset: the standard deviation, $\sigma_{\overline{\theta}}$, of the mean of the obtained angles.}
    \label{obtained vs unknown angles, N=20}.    
\end{figure*}

\subsection{Phase Sensing} 
The SN can also be used as a phase sensor device, that retrieves an unknown phase applied at a known site and time. Assuming an unknown phase factor of $e^{i\theta}$ is applied at site $\frac{N}{2}+1$ at $t=t_m$, the task is to retrieve this unknown phase angle, $\theta$. In our previous work on the SN of size $N=6$, \cite{alsulami2022unitary}, we have proposed a protocol that can be used to retrieve this unknown angle even in the presence of significant error. Using the proposed phase sensing protocol, we investigate below the robustness of the phase sensing for larger SN systems.

\subsubsection{Phase Sensing Robustness} 
The robustness of the phase sensing protocol has been investigated against coupling disorder and on-site energy disorder with relatively large error strengths of $E=5\%$ and $E=10\%$. The performance of our phase sensing protocol in retrieving a range of unknown angles ranging from $0\si{\degree}$ to $360\si{\degree}$ (in degrees), for different SN sizes, is illustrated in Fig.\eqref{obtained vs unknown angles, N=20}.

It is clear from Fig.\eqref{obtained vs unknown angles, N=20} that the obtained angles in the presence of on-site energy disorder are very similar to the ideal case, which is attributed to the strong robustness of the SN in the presence of diagonal disorder. In the coupling disorder case with $E=5\%$, Fig.\eqref{obtained vs unknown angles, N=20}(a,c), the obtained angles are almost accurate except for angles around 45\si{\degree} where they deviate slightly from the ideal case, especially for $N=50$. On the other hand, the obtained angles in the presence of coupling disorder with $E=10\%$, Fig.\eqref{obtained vs unknown angles, N=20}(b,d), shows deviation from the ideal case for various unknown angles, especially for large SN of $N=50$. These observation can also be understood by looking at the mean of the standard deviation, $\sigma_{\overline{\theta}}$, which clearly shows that there is clear fluctuation in the coupling disorder case verses small fluctuation in the on-site energy disorder case.  

Therefore, our SN device can be scaled up to $N=50$ and still be used as a phase sensor with very good performance against both types of disorder, as long as the error strength is $E\leq5\%$. This good performance can still be observed when $E=10\%$, even for $N=50$ if the errors in the system are due to on-site energy disorder, and for $N=20$ if the errors arise from coupling disorder. As already discussed in Section \ref{2_equal_ch_SN_router_robustness}, these figures demonstrate that the scalability (increasing $N$) has a smaller impact on the protocol, compared to increasing the amount of error strength $E$. 

\subsection{Spin Networks of unequal chains}
\label{Unequal-chain SN}
\begin{figure}[ht!]
    \centering
    \includegraphics[width=0.5\textwidth]{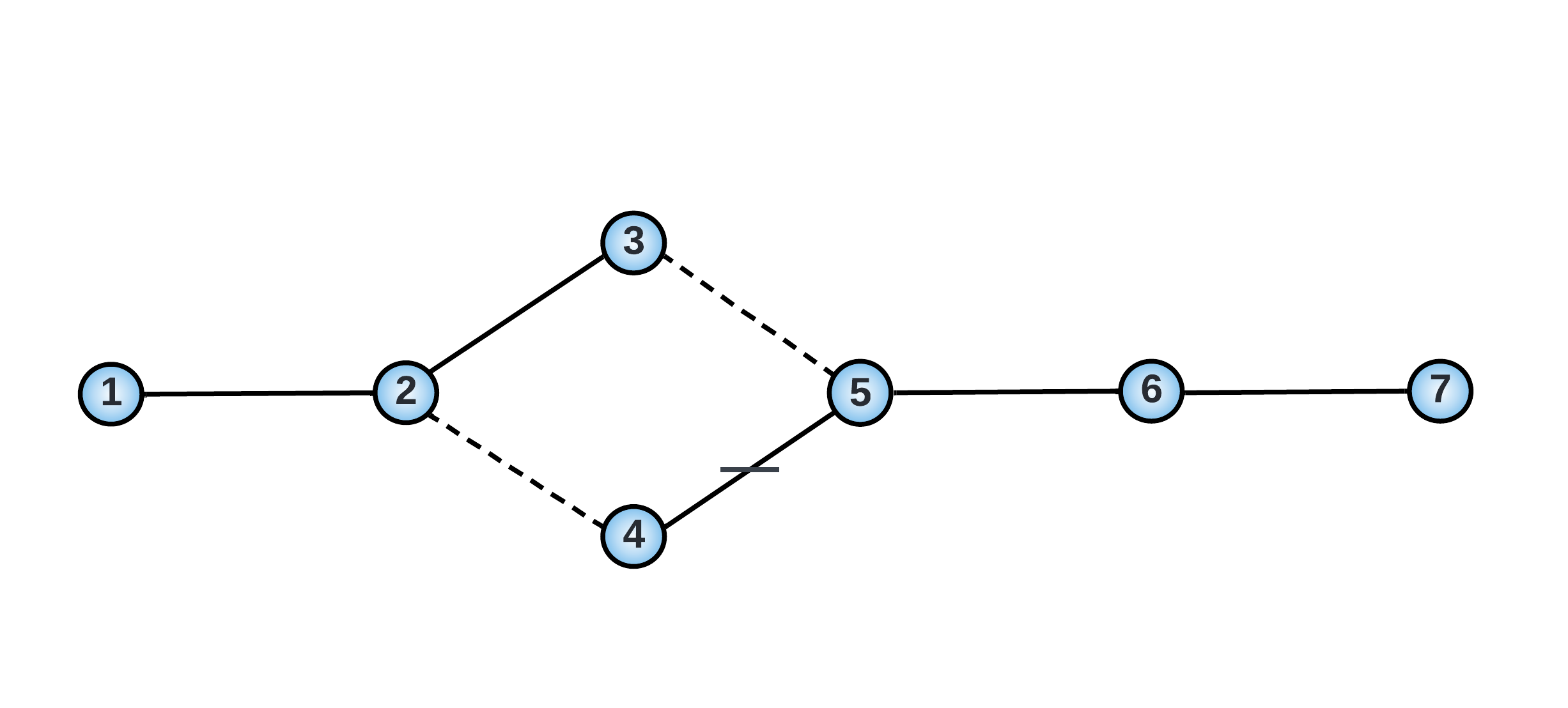}
    \caption{Example of a SN made by coupling two unequal PST chains, in this case a 3-site PST chain with a 4-site PST chain.}
    \label{SN of 7sites}.
\end{figure}

We now investigate a SN system of unequal chains (i.e., SN designed by coupling together two PST chains of different length). For example, a SN of two PST chains, $A$ and $B$ of $N_A=3$ sites and $N_B=4$ sites, respectively, is shown in Fig.\eqref{SN of 7sites}. Since the chains in this SN are not equal, the time evolution through each chain is different. Therefore, we denote the mirroring time of the first chain and the second chain as $t_{m,A}$ and $t_{m,B}$, respectively. This SN can be used too to generate the routing and entanglement protocols, as discussed below.

\subsubsection{Routing}
The routing protocol, in the error-free case, is achieved in the SN regardless of the SN size and regardless of whether the SN chains are equal or not. For example, if in the SN shown in Fig.\eqref{SN of 7sites}, we inject a single-excitation at site 1 at $t=0$ and a phase flip at site 3 at $t=t_{m,A}$, then after another evolution of $t=t_{m,B}$ the state of the system will be given by
\begin{equation}
\label{SN of unequal chains router}
\ket{\psi(t_{m,A,B})} = -i\ket{r_7}\; 
\end{equation}
where $t_{m,A,B}=t_{m,A}+t_{m,B}$ denotes the time the excitation needed to evolve from site 1 to site 7.

In the general case, for a longer 2-chains SN of arbitrary length $N_A+N_B=N$, the routing protocol will still work in the same way and the routed state will be given by  
\begin{equation}
\label{SN of unequal chains router general case}
\ket{\psi(t_{m,A,B})} = e^{-i\gamma(N)}\ket{r_N}\; 
\end{equation}

\subsubsection{Entanglement generation}
Generation of a bipartite maximally entangled state between the ends of this SN is not straightforward due to the different lengths of the SN chains. 

When a single-excitation is injected at site 1 at $t=0$ and a phase of $e^{i\pi/2}$ is injected at site 4 at $t_{m,A}$, amplitudes of the excitation will then propagate through the first chain (the shorter one) and through the second chain (the longer one). At $2t_{m,A}$, the amplitude of the excitation that evolves through the first chain will be localized at site 1, while the excitation's amplitude that evolves through the second chain will be delocalised over its sites. Therefore, the state of the system at $2t_{m,A}$ will be given by 
  
\begin{equation}
\label{Entang fid} 
\begin{split}
    \ket{\psi(2t_{m,A})} &= \frac{1+e^{i\pi/2}}{2}\ket{r_1} 
    + a(\ket{r_3}-\ket{r_4}) \\
    &+ b\ket{r_5} + c\ket{r_6} + f\ket{r_7}\; 
\end{split}        
\end{equation}
where $a\approx-0.031+0.031i$, $b\approx0.15+0.15i$, $c\approx0.31-0.31i$, and $f\approx-0.36-0.36i$.

\begin{figure}[H]
    \centering
    \includegraphics[width=0.5\textwidth]{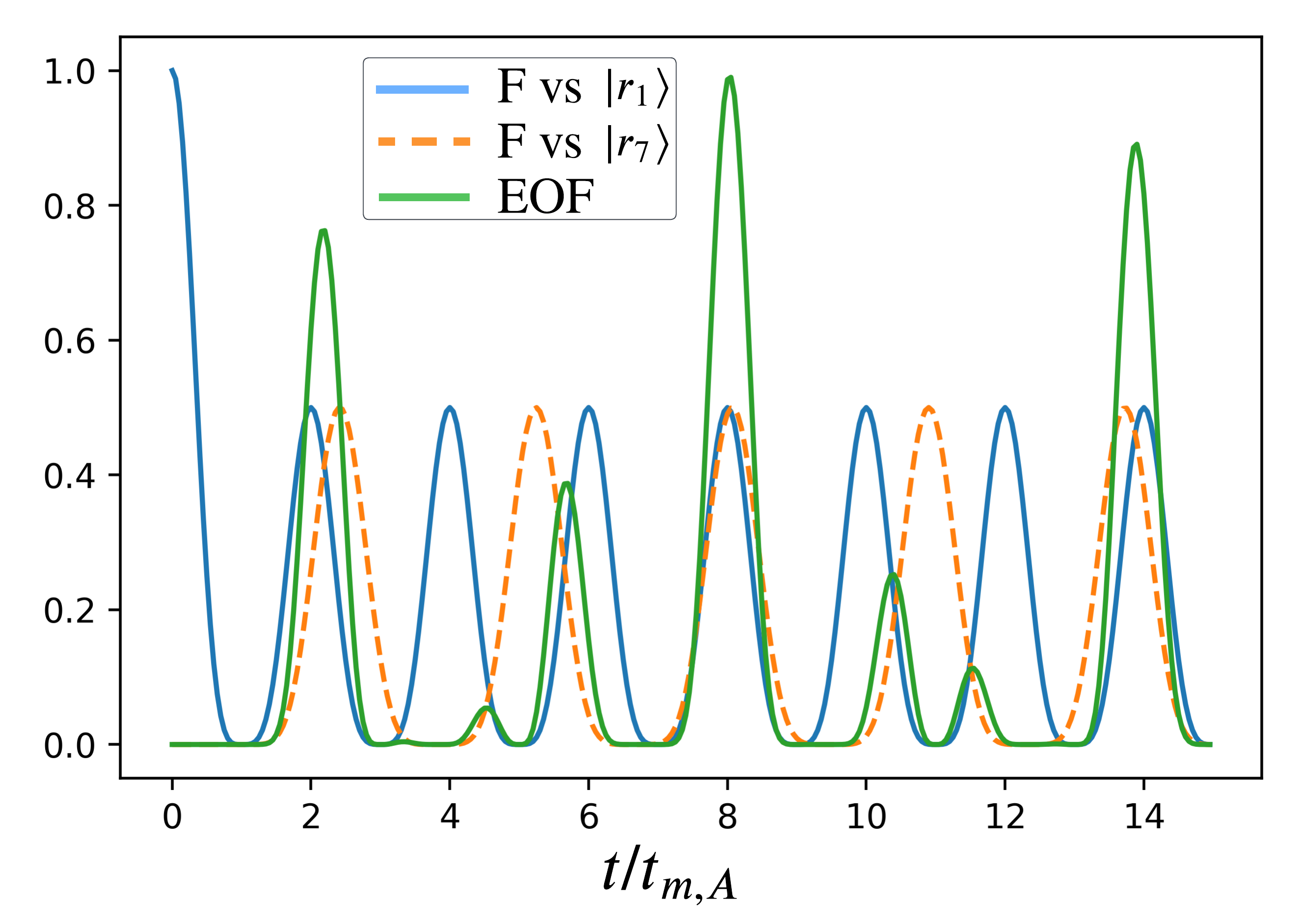}
    \caption{Fidelities of the system against desirable state $\ket{r_1}$ (F vs $\ket{r_1}$, blue) and against desirable state $\ket{r_7}$ (F vs $\ket{r_7}$, orange) and the EOF between them, green, as labelled in figure.}%
    \label{Fid vs r_1 & vs r_7 and EOF of r1andr4}
\end{figure}

The fidelity of the system against $\ket{r_1}$, against $\ket{r_7}$, and the EOF between sites 1 and 7 are plotted as a function of time in Fig.\eqref{Fid vs r_1 & vs r_7 and EOF of r1andr4}. The plot demonstrates that, at $2t_{m,A}$, half of the excitation is at site 1 and the other half is in a superposition as given in Eq.\eqref{Entang fid}. It is also clear from the plot that after few oscillation, at $t=8t_{m,A}$, the state of the system is almost a maximally entangled state between sites 1 and 7 as the EOF is very close to 1. This however is not perfect generation of the desired state, as the mirroring times of the two chains are not equal. This  can be resolved by exploiting the dependence of $t_m$ on $J_{max}$ (see Section \ref{model sec}) and adjusting the maximum coupling of one of the chains.

If we choose to adjust the maximum coupling of chain $B$ (which is the longer chain), this will result in an increase of its maximum coupling $J_{max,B}$, thus speeding up its time evolution to match that of chain $A$ ($t_{m,B}=t_{m,A}$). This adjustment of the $J_{max,B}$ is given by
\begin{equation}
   J_{max,B}=\frac{\pi N_B}{4t_{m,A}}, \textcolor{white}{aa}\text{if} \textcolor{white}{aa} N_B \textcolor{white}{aa} \text{is even};
\end{equation}
\begin{equation}
   J_{max,B}=\frac{\pi\sqrt{\frac{N_B^2-1}{4}}}{2t_{m,A}}, \textcolor{white}{aa}\text{if} \textcolor{white}{aa} N_B \textcolor{white}{aa} \text{is odd};
\end{equation}


If instead we adjust the maximum coupling of chain $A$ (which is the shorter chain), this will result in a decrease of its maximum coupling $J_{max,A}$, thus retarding its time evolution to match that of chain $B$ ($t_{m,A}=t_{m,B}$). This adjustment of the $J_{max,A}$ is given by 
\begin{equation}
   J_{max,A}=\frac{\pi N_A}{4t_{m,B}}, \textcolor{white}{aa}\text{if} \textcolor{white}{aa} N_A \textcolor{white}{aa} \text{is even};
\end{equation}
\begin{equation}
   J_{max,A}=\frac{\pi\sqrt{\frac{N_A^2-1}{4}}}{2t_{m,B}}, \textcolor{white}{aa}\text{if} \textcolor{white}{aa} N_A \textcolor{white}{aa} \text{is odd};
\end{equation}


Both solutions (adjusting the shorter or the longer chain) are viable solutions where physically possible, but we choose to adjust the shorter chain, $A$, which relies on reducing the coupling strengths. This is because adjusting the longer chain, $B$, requires increasing couplings, which might not be practical. Specifically, when $N_B\gg N_A$, if an experimental constraint on the maximum coupling is already saturated in the uncoupled chain $B$, no further increase of these couplings would be possible.

Having set the mirroring time of both chains to be equal, we can now generate a bipartite maximally entangled state between the ends of the SN, following the protocol given in Fig.\eqref{Entanglement protocol} or the second entanglement protocol given in  section \ref{seconed entang protocol}. Let us use the second protocol, where we start with a single-excitation at site 3 at $t=0$ and let the system evolve for a duration of $t_{m,A}$. The state at $t_{m,A}$ will then be given as a bipartite maximally entangled state between sites 1 and 7  
\begin{equation}
\label{entang state SN of unequal ch_N=7}
\ket{\psi(t_{m,A})} = \frac{1}{\sqrt{2}}(-\ket{r_1}+i\ket{r_7}). 
\end{equation} For a SN of unequal chains of an arbitrary length, injecting a single-excitation at the top site of the central vertex of the SN diamond (e.g., site 3 in Fig.\ref{SN of 7sites}) at $t=0$ will naturally generate a bipartite maximally entangled state between sites 1 and $N$ at $t_{m,A}$ given as
\begin{equation}
\label{entang state SN of unequal ch}
\ket{\Psi(t_{m,A})} = \frac{1}{\sqrt{2}}(e^{-i\alpha(N_A)}\ket{r_1}+e^{-i\alpha(N_B)}\ket{r_N})\; ,
\end{equation}
where the phase factors $e^{-i\alpha(N_j)}=(-i)^{N_j-1}$, $j=A,~B$.

\begin{figure}[H]
    \centering
    \includegraphics[width=0.5\textwidth]{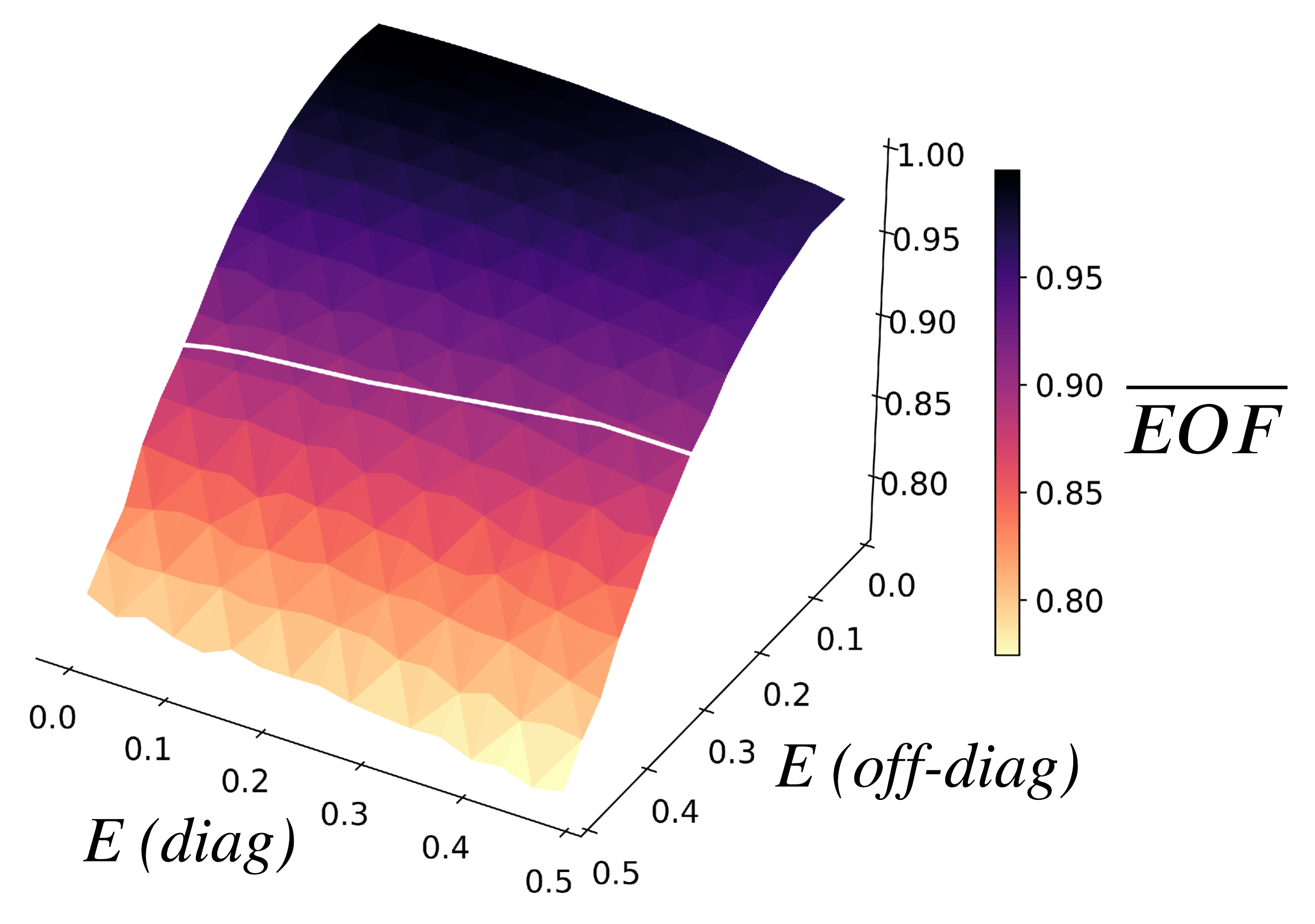}
    \caption{The robustness of the $\overline{EOF}$ of the bipartite maximally entangled state (generated using the second entanglement protocol) at the first time it forms, $t_{m,A}$, in the presence of diagonal disorder (\textit{diag}) and off-diagonal disorder (\textit{off-diag}) with different error strengths, $E$, ranging from 0 to $50\%$. Each point has been averaged over 1000 realizations . White line has same meaning as in Fig.~\ref{EOF robustness vs N and vs disorders}}.
    \label{bipartite_vs_errors_SN_of_3s_and_4s_SC}
\end{figure}

The robustness of the bipartite maximally entangled state generated in Eq. \eqref{entang state SN of unequal ch_N=7} is investigated by calculating the $\overline{EOF}$ between sites 1 and 7 at the first time it forms ($t_{m,A}$). This is shown in Fig.\ref{bipartite_vs_errors_SN_of_3s_and_4s_SC} where the robustness of $\overline{EOF}$ in the presence of diagonal disorder with significant error strength of $E=20\%$ is $\approx99.5\%$. In the presence of off-diagonal disorder, the $\overline{EOF}$ is found to be $>97\%$ for error strength of $E\leq10\%$.

\section{Multiple-chain Spin Networks}
\label{Multi-chains SN}
Thus far, we have investigated a SN that is built by coupling together two chains. A more general method of scaling the SN system is to couple together more than two spin chains. The concept of modularisation (i.e., connecting together multiple identical systems) has been used for high-fidelity QST \cite{almeida2016quantum} and modular entanglement \cite{gualdi2011modular}. We now move to the discussion of SN with more than two component spin chains. The building blocks of these multiple-chain SN are once more uncoupled PST chains, coupled together by unitary transformation, \cite{alsulami2022unitary}.

\subsection{Example: Three chains of 3-sites SN}
\label{example A}
\begin{figure}[H]  
    \centering
    \includegraphics[width=0.45\textwidth]{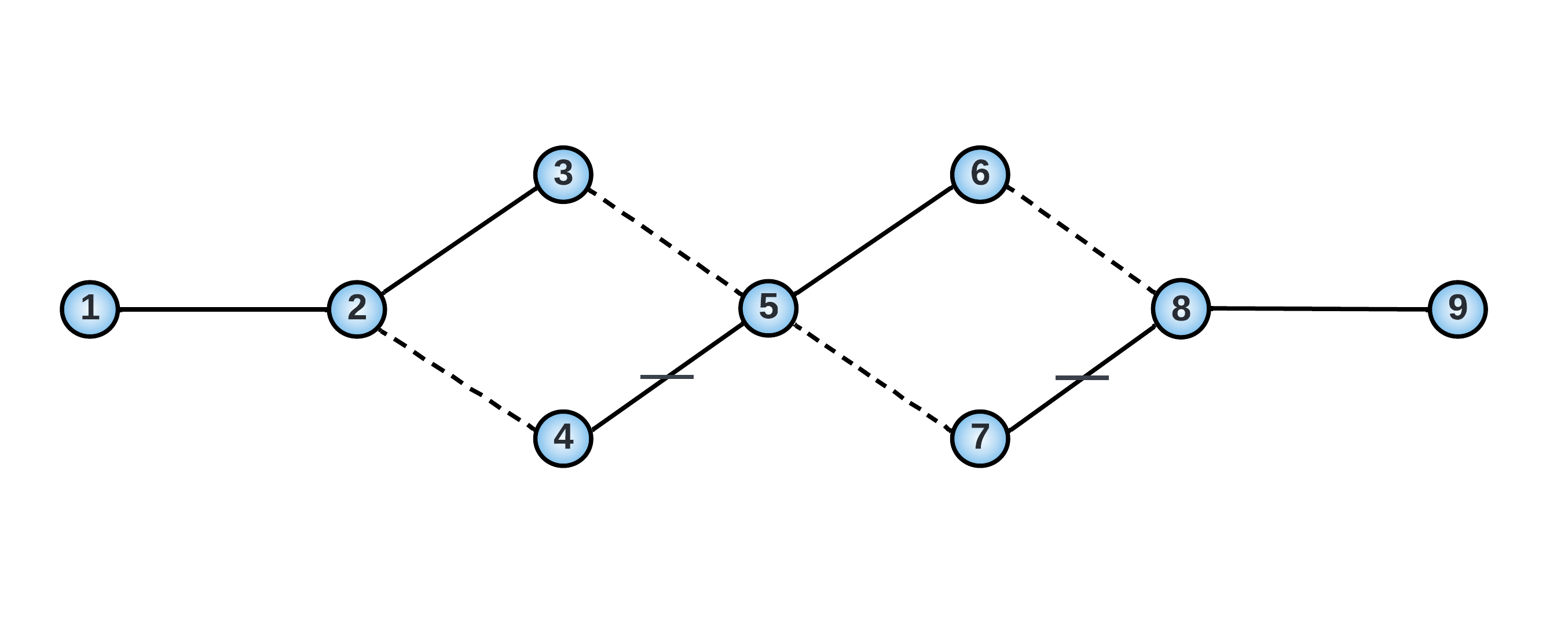}
    \caption{A 3-chain SN, each of 3-sites. Sites 1, 2, and 3 represent the first chain. Sites 4, 5, and 6 represent the second chain. Sites 7, 8, and 9 represent the third chain.} 
    \label{Ideal 9s SN}.     
\end{figure}

The SN we discuss here is built by coupling together 3 PST chains, each of 3-sites, Fig.\eqref{Ideal 9s SN}. The unitary transformation used to transform the Hamiltonian of the uncoupled chains is chosen such that it superposes sites 3 and 4 as well as sites 6 and 7. The matrix of the unitary transformation for the single-excitation basis is given as  

\begin{equation}
\label{equationSN H} 
\begin{split}
 U = \begin{pmatrix}       
    1&0&0&0&0&0&0&0&0 \\
    0&1&0&0&0&0&0&0&0 \\
    0&0&\frac{1}{\sqrt{2}}&\frac{1}{\sqrt{2}}&0&0&0&0&0 \\
    0&0&\frac{1}{\sqrt{2}}&\frac{-1}{\sqrt{2}}&0&0&0&0&0 \\
    0&0&0&0&1&0&0&0&0 \\
    0&0&0&0&0&\frac{1}{\sqrt{2}}&\frac{1}{\sqrt{2}}&0&0 \\
    0&0&0&0&0&\frac{1}{\sqrt{2}}&\frac{-1}{\sqrt{2}}&0&0 \\
    0&0&0&0&0&0&0&1&0 \\
    0&0&0&0&0&0&0&0&1 \\
\end{pmatrix}
\end{split}
\end{equation} \;

Our SN indeed has the potential to achieve various interesting phenomena and therefore it can be used to support various tasks in quantum technologies, including generating different types of entangled states, as will be seen below.

\subsubsection{Routing protocol} The routing protocol is straightforward and can be achieved by injecting a single-excitation at site 1 at $t=0$, a phase flip at site 4 at $t=t_m$, and a another phase flip at site 7 at $t=2t_m$, which will then result in an excitation being at site 9 at $t=3t_m$.    

\subsubsection{W-State Entanglement} 
The entanglement based on an excitation shared equally between three sites is generally called W-state entanglement \cite{guhne2009entanglement,m2019tripartite}. This kind of entanglement not only provides the possibilities to investigate quantum non locality \cite{ZhangChao2016Etog} but also have various applications in quantum information protocols such as quantum teleportation \cite{shi2002teleportation}, superdense coding \cite{zhou2018efficientdense}, and quantum secure direct communication \cite{chen2008controlled}. Our SN, Fig.\eqref{Ideal 9s SN}, can be used to generate W-state entanglement as will be discussed below. 

When a single-excitation is injected at site 1 at $t=0$, it will evolve to be in a superposition state being at sites 3 and 4 at $t=t_m$. If at this time an arbitrary phase of $e^{i\theta}$ is applied at site 4, the resultant evolved state at $t=2t_m$ will be given by  
\begin{equation}
    \ket{\psi(2t_m)} = \frac{1+e^{i\theta}}{2}\ket{r_1}
    +\frac{1-e^{i\theta}}{2\sqrt{2}}(\ket{r_6}+\ket{r_7}).
\end{equation}
Therefore, if the phase factor applied at site 4 at $t_m$ is $e^{i\phi}$, where $\phi=\arccos(-1/3)$, then the resultant evolved state at $t=2t_m$ will be an equal superposition state between the excitation being at sites 1, 6, and 7, which is a W-state
\begin{equation}
    \ket{W} = \frac{1+e^{i\phi}}{2}\ket{r_1}
    +\frac{1-e^{i\phi}}{2\sqrt{2}}(\ket{r_6}+\ket{r_7}).
\end{equation}

\begin{figure}[H]
    \centering
    \includegraphics[width=0.45\textwidth]{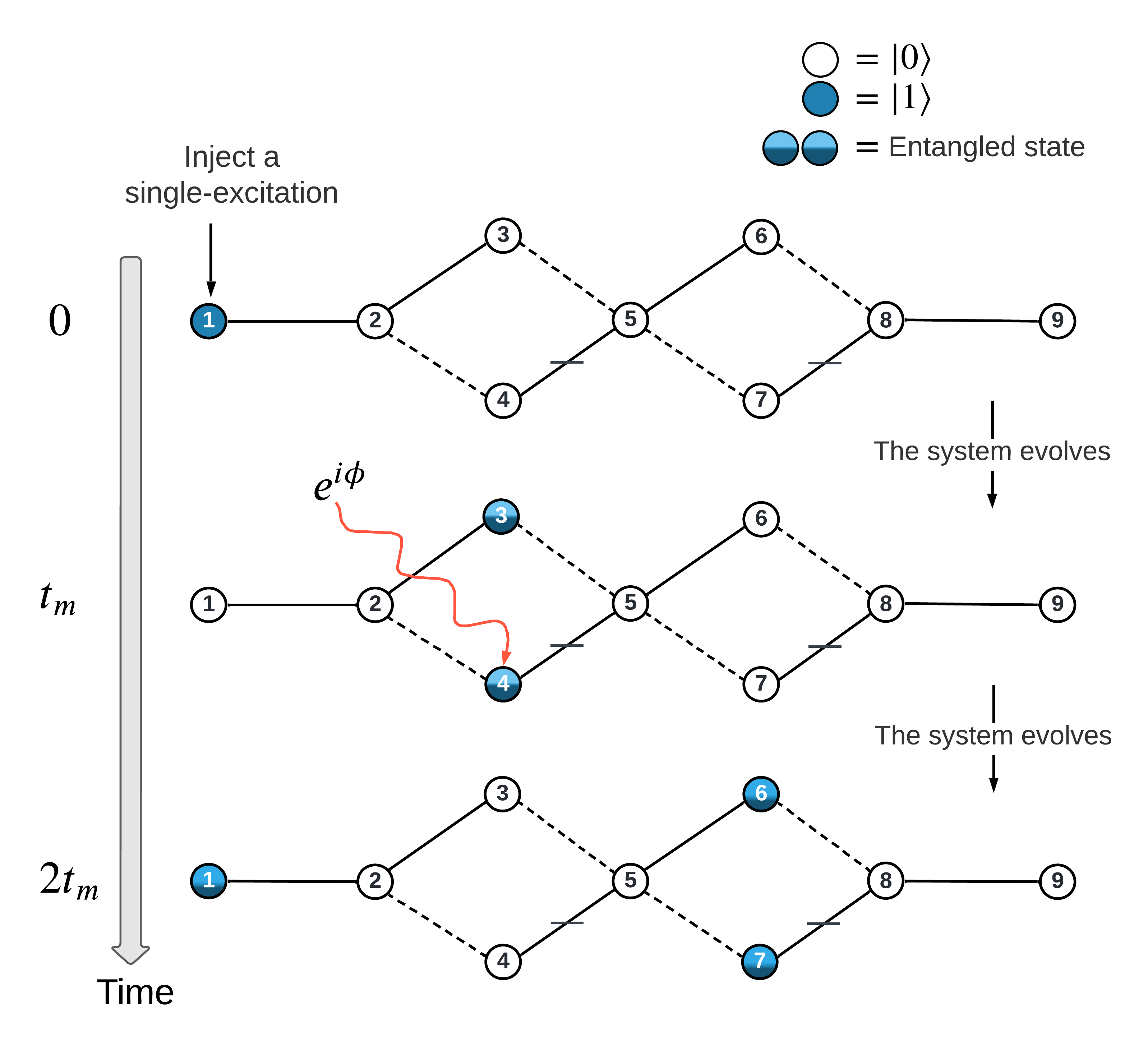}
    \caption{Generation of a W-state entanglement. The phase applied at site 4 is $\phi=cos^{-1}(-1/3)$.} 
    \label{W entanglement protocol in the 9s SN}.     
\end{figure}

Due to the periodicity of our SN, if the system is left to evolve, then the W state will keep forming at regular time intervals as shown in Fig.\eqref{Fids of each site, W-state, n=9}. The protocol for generating the W-state entanglement is demonstrated in Fig.\eqref{W entanglement protocol in the 9s SN}. 

\begin{figure}[H]
    \centering
    \includegraphics[width=0.45\textwidth]{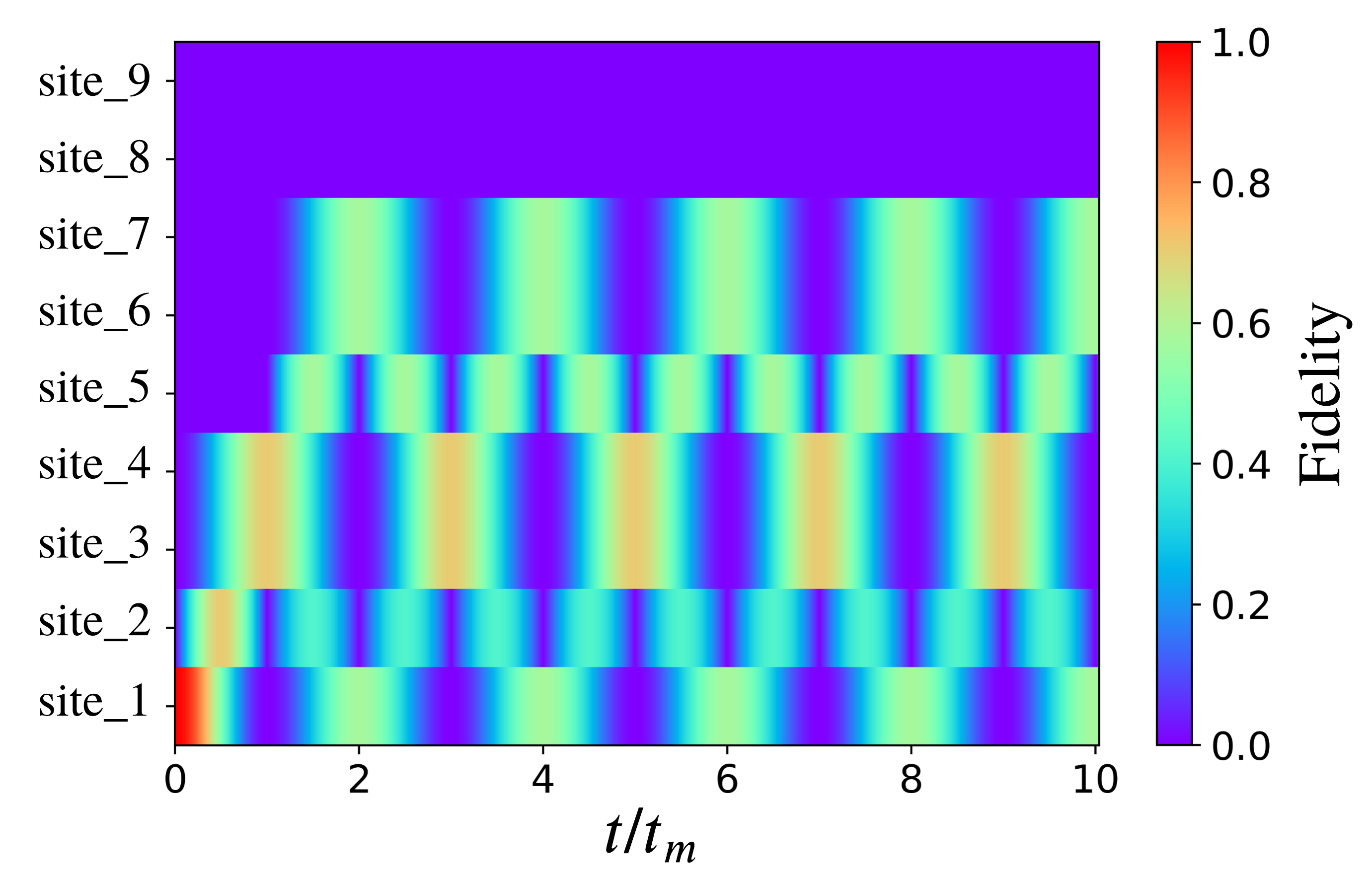}
    \caption{Fidelity of each site as a function of the rescaled time $t/t_m$ shows that the W state generated between sites 1, 6, and 7 keeps forming at each even $t_m$ (i.e., $2t_m$, $4t_m$, $6t_m$, $\ldots$).} 
    \label{Fids of each site, W-state, n=9}.     
\end{figure} 

\begin{figure}[ht!]
    \centering
    \includegraphics[width=0.45\textwidth]{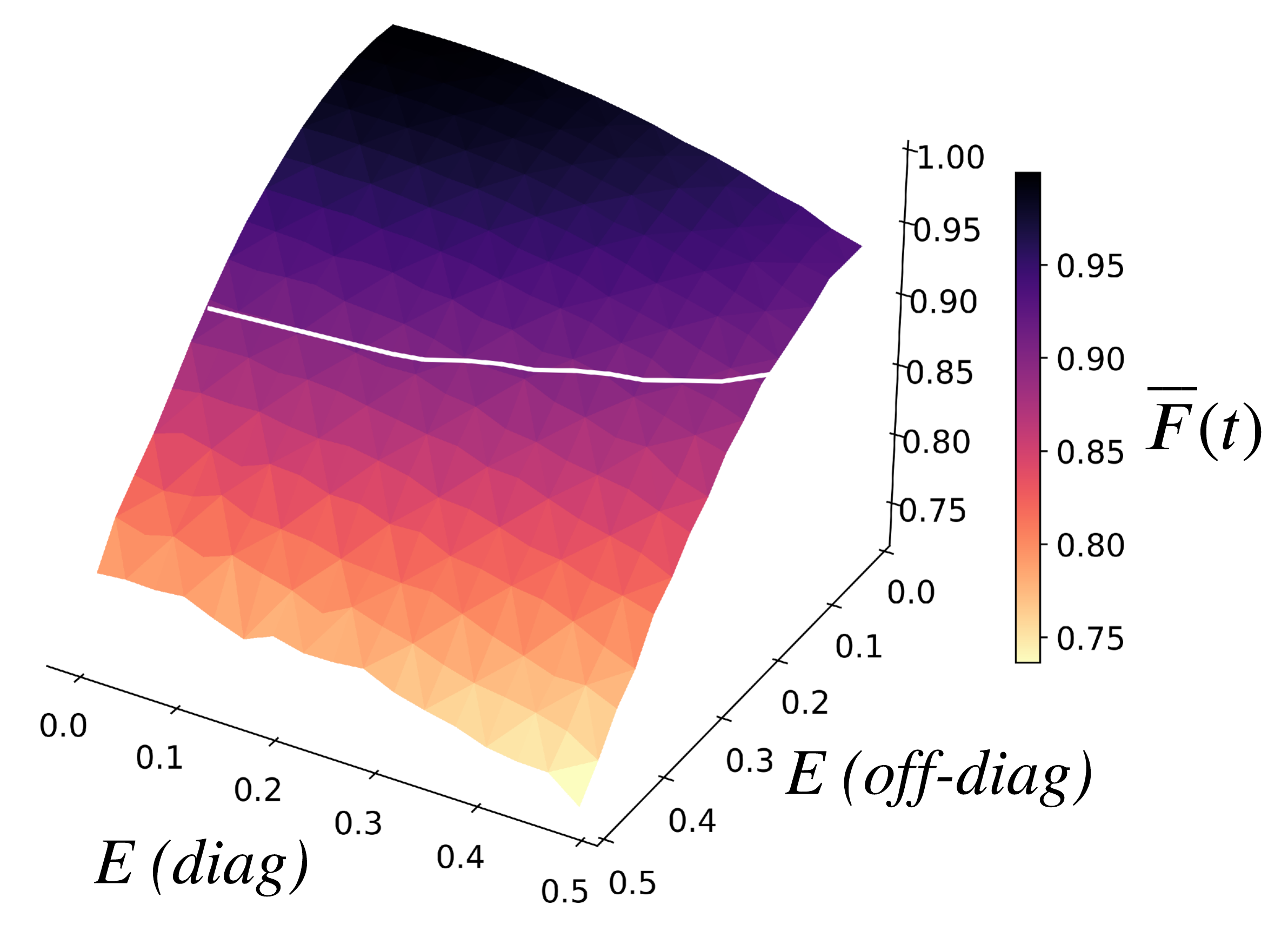}
    \caption{The robustness of the fidelity of the W entangled state at the first time it forms, $2t_m$, in the presence of diagonal disorder (\textit{diag}) and off-diagonal disorder (\textit{off-diag}) with different error strengths, $E$, ranging from 0 to $50\%$. Each point has been averaged over 1000 realisations. White lines have same meaning as in Fig.~\ref{router robustness vs N and vs disorders}.
    }%
    \label{W entanlged robustness vs err}.     
\end{figure}

The robustness of the W-state entanglement generated in Fig.\eqref{W entanglement protocol in the 9s SN} is investigated by measuring the fidelity of the system with a desirable state chosen as a W entangled state. The fidelity is therefore measured at the first time the W entangled state is expected ($2t_m$) in the presence of both types of disorder, Fig\eqref{W entanlged robustness vs err}.

The W-state entanglement turns out to be very robust against diagonal disorder, giving fidelity up to $97\%$ with a significant error strength of $E\leq25\%$ and fidelity of $98.5\%$ with $E\leq5\%$. In the presence of off-diagonal disorder, the W-state robustness still very high with fidelity $\geq95\%$ against significant error strength of $E\leq20\%$ and decays faster as the error strength increases.

\subsubsection{Other Entangled States}
\label{other entang states}
Entanglement based on an excitation shared equally between four or more sites is called Multipartite W-type State (MWS) entanglement \cite{walter2016multipartite}. We will now show how a MWS entanglement (shared between four sites) can be generated. 

The system is first initialised with a single-excitation injected at site 5 at $t=0$
\begin{equation}
\label{inis5}
    \ket{\psi(0)}=\ket{r_5} 
\end{equation}
and let to evolve for time $t=t_m/2$. The resulted evolved state will be given by an equal superposition state between four sites, a MWS entanglement
\begin{equation}
\label{MWS5} 
    \ket{\psi(t_m/2)}=\frac{i}{2}(-\ket{r_3}+\ket{r_4})-\frac{i}{2}(\ket{r_6}+\ket{r_7})\;.
\end{equation}

If we do nothing, the excitation will then evolve back to site 5 at $t=t_m$ and keeps evolving between Eq.\eqref{inis5} and Eq.\eqref{MWS5}, due to the periodicity of the system. However, as we have seen above, the application of a local phase can change the direction of the excitation evolution. Therefore, we will use this to generate a bipartite maximally entangled state (between the ends of the SN). 

\begin{figure}[ht!]
    \centering
    \includegraphics[width=0.45\textwidth]{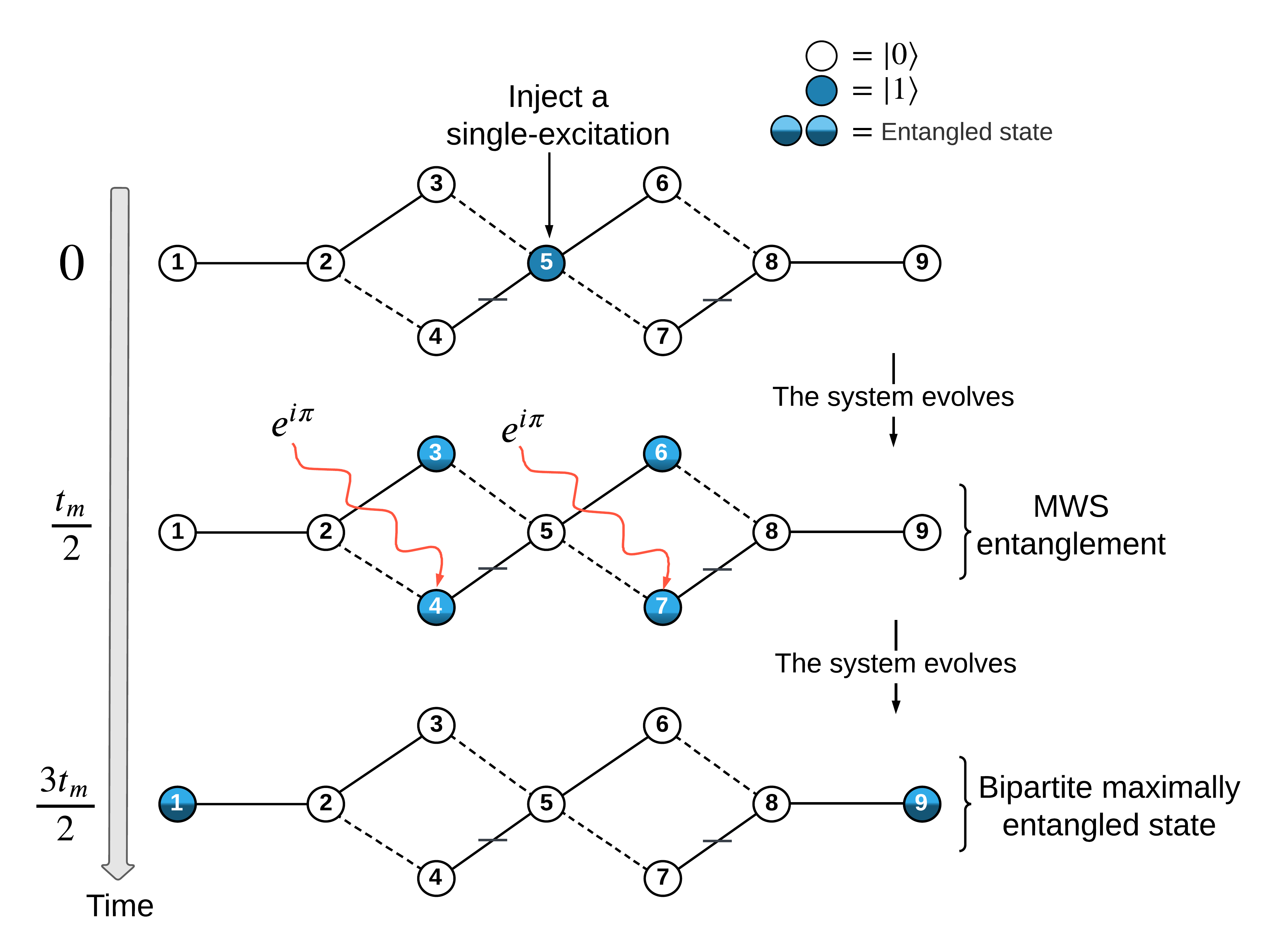}
    \caption{Illustration of the injection protocol used to obtain a MWS entanglement between sites 3, 4, 6, and 7. Additional two phase flip can be applied simultaneously at sites 4 and 7 in order to obtain a bipartite maximally entangled state between the ends of the SN. The excitation then keeps evolving between a MWS entanglement and a bipartite maximally entangled state.}  
    \label{Multipartite and Maximally entangled protocol}.     
\end{figure}

A bipartite maximally entangled state between the ends of the SN (sites 1 and 9 in Fig.\eqref{Ideal 9s SN}) can be generated by injecting two phase flip simultaneously at sites 4 and 7 at $t_m/2$, the time the system state is in a MWS entanglement, Eq.\eqref{MWS5}. Then, the system is left to evolve for a further duration of $t_m$ and will result in the desired state given by 
\begin{equation}
    \ket{\psi(3t_m/2)}=\frac{i}{\sqrt{2}}(\ket{r_1}+\ket{r_9})\;.
\end{equation}
The generation of these two types of entangled states is illustrated in Fig.\eqref{Multipartite and Maximally entangled protocol}.  

\begin{figure}[H]
    \centering
    \includegraphics[width=0.45\textwidth]{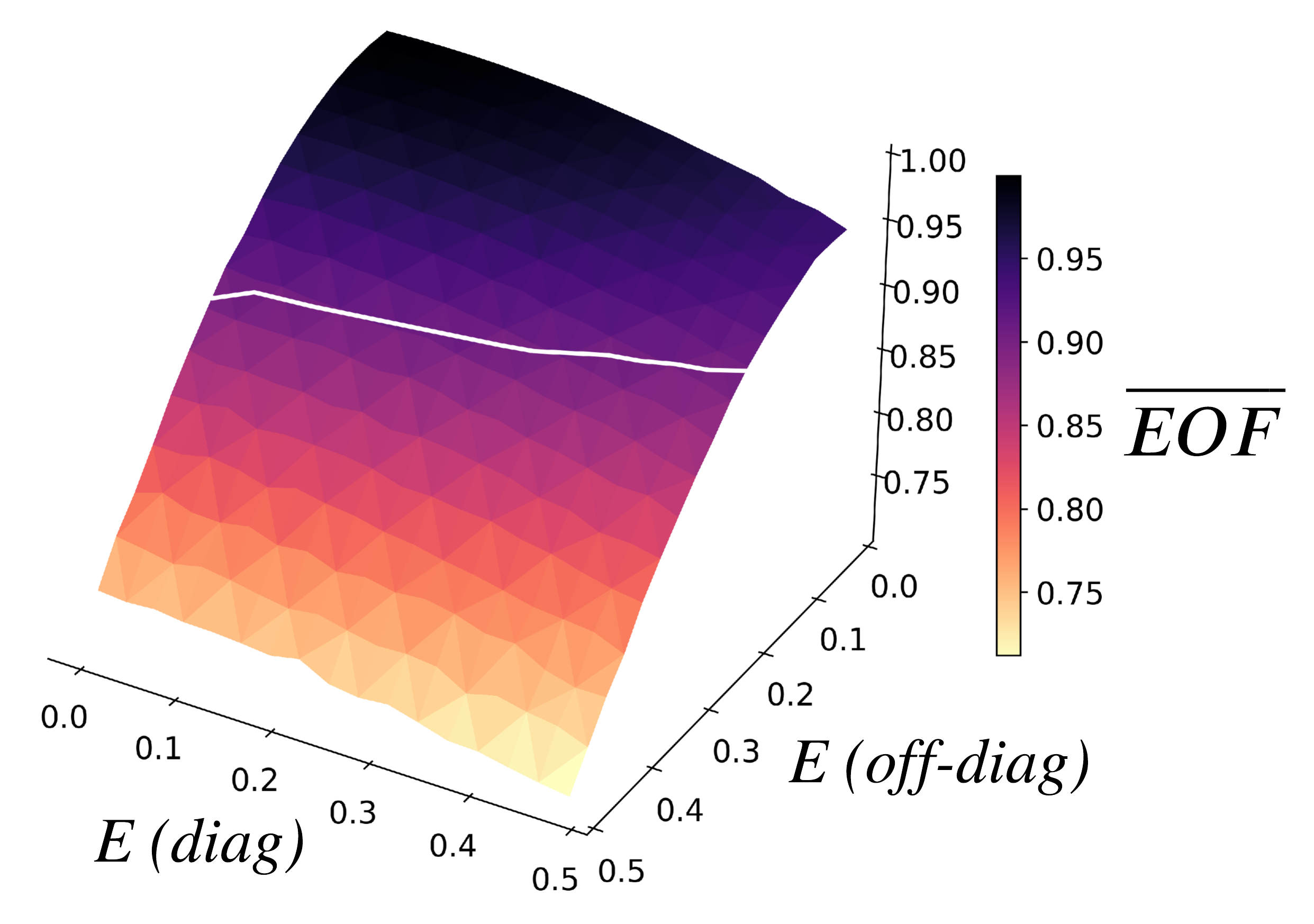}
    \caption{The robustness of the $\overline{EOF}$ of the bipartite maximally entangled state at the first time it forms, $3t_m/2$, in the presence of diagonal disorder (\textit{diag}) and off-diagonal disorder (\textit{off-diag}) with different error strengths, $E$, ranging from 0 to $50\%$. Each point has been averaged over 1000 realizations . White line has same meaning as in Fig.~\ref{EOF robustness vs N and vs disorders}.
    }
    \label{Maximally robustness vs err}.     
\end{figure}

The robustness of the bipartite maximally entangled state generated in Fig.\eqref{Multipartite and Maximally entangled protocol} is investigated by measuring the $\overline{EOF}$ between sites 1 and 9 at the first time it forms ($3t_m/2$), against both types of disorder, Fig\eqref{Maximally robustness vs err}. The bipartite maximally entangled state is also very robust, giving fidelity up to $99\%$ against diagonal disorder with large error strength of $E\leq15\%$. Whereas in the presence of off-diagonal disorder, its fidelity scales up to $98\%$ with error strength of $E\leq10\%$. The robustness of our SN system in generating bipartite maximally entangled state gives our SN the potential to be used in various quantum information processing. 

\subsection{Example: Three chains of 4-sites SN}
\begin{figure*}
    \includegraphics[width=0.8\textwidth]{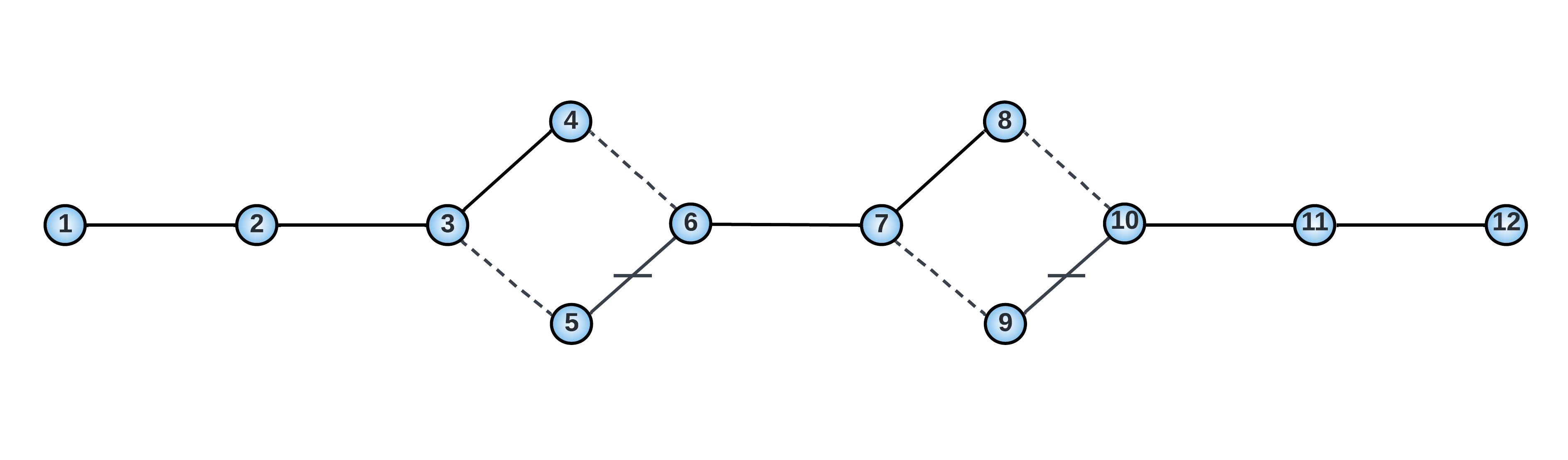}
    \caption{Diagram of a 3-chain SN, each of 4-sites.} 
    \label{fig:20}.     
\end{figure*}

Let's now extend our investigation to a larger SN, that is built by coupling together 3 PST chains, each of 4-sites, Fig.\eqref{fig:20}. The unitary transformation used to transform the Hamiltonian of the uncoupled chains is chosen such that it superposes sites 4 and 5 as well as sites 8 and 9.

\subsubsection{W-State Entanglement}
Generation of a W-state entanglement is also possible in this SN, as described below. 

By injecting a single-excitation at site 1 at $t=0$ and a phase factor of $e^{i\phi}$, where $\phi=\arccos(-1/3)$, at site 5 at $t=t_m$, the resultant evolved state at $t=2t_m$ will be an equal superposition state between sites 1, 8, and 9  
\begin{equation}
    \ket{W} = -\frac{1+e^{i\phi}}{2}\ket{r_1}
    -\frac{1-e^{i\phi}}{2\sqrt{2}}(\ket{r_8}+\ket{r_9}).
\end{equation}

\begin{figure}[ht!] 
    \includegraphics[width=0.5\textwidth]{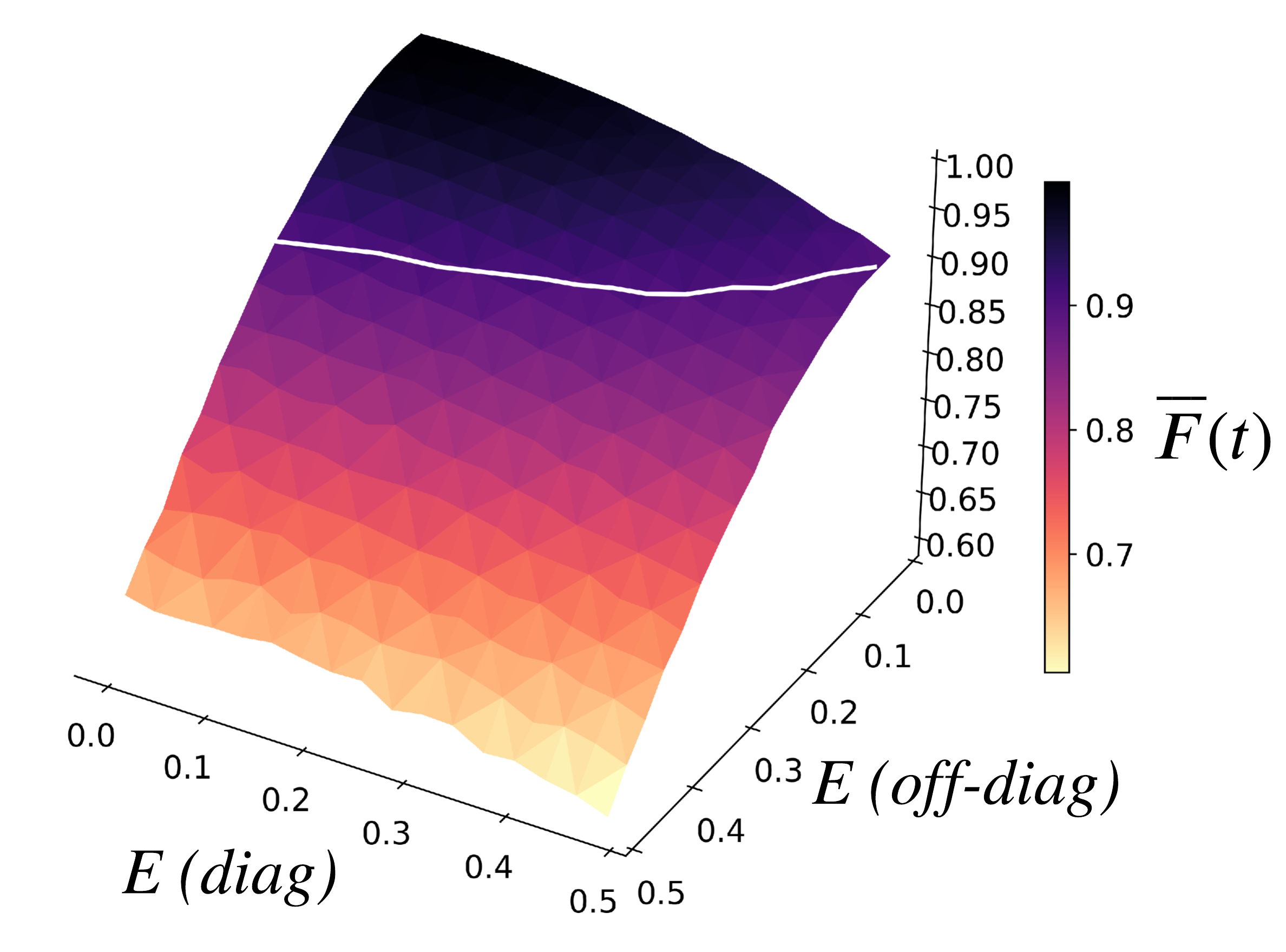} 
    \caption{Demonstration of the W-state robustness in the presence of diagonal disorder (\textit{diag}) and off-diagonal disorder (\textit{off-diag}) with different error strengths, $E$, ranging from 0 to $50\%$. Each point has been averaged over 1000 realisations. White line has same meaning as in Fig.~\ref{router robustness vs N and vs disorders}.
    }%
    \label{w-state robustness diag_err}%
\end{figure}

The robustness of the W-state against different types of disorder is shown in Fig.\eqref{w-state robustness diag_err}. In the presence of diagonal disorder, the W-state robustness remains $\stackrel{>}{\sim}99\%$ up to $E=15\%$ when it is measured at $t=2t_m$. On the other hand, the W-state robustness against off-diagonal disorder remains $\stackrel{>}{\sim}98\%$ for $E\leq10\%$ and decays as the error strength increases, $E\geq10\%$. Overall, the W-state is very robust as long as $E\leq10\%$, which suggests the potential of our SN for practical W-state entanglement generation. 

\subsubsection{MWS Entanglement}
\label{Sec. MWS entang}
We have seen in the previous example, Section \ref{other entang states}, that it was straightforward to generate a MWS entanglement due to the fact that there exist a site (i.e., site 5 in Fig.\eqref{Ideal 9s SN}) that is directly coupled with the central vertices of the diamonds. However, this is not the case in the SN shown in Fig.\eqref{fig:20}. Therefore, we will use another protocol to generate the MWS entanglement in this SN. 

In our SN, Fig.\eqref{fig:20}, we seek to generate a MWS entanglement between sites 4, 5, 8, and 9. In order to do so, the SN coupling parameters need to be adjusted such that the excitation evolution time through the central chain is not equal to the excitation evolution time through the other chains. This is done by setting the maximum coupling of the second chain to be $J_{max,B}=1/2$, whilst keeping the maximum coupling of the first and third chains to be $J_{max,A}=J_{max,C}=1$. The letters $A$, $B$, and $C$ represent the first, second, and third chains of the SN. Consequently, the mirroring time for the second chain will be equal to twice the mirroring time of the first chain (i.e., $t_{m,B}=2t_{m,A}$). 

Having established that, we now start by initialising the system with a single-excitation injected at site 5 at $t=0$
\begin{equation}
    \ket{\psi(0)}=\ket{r_5}.
\end{equation}
A MWS entanglement can then be generated by evolving the initial state for time $t=2t_{m,A}$. Over this period of time, half of the excitation amplitude evolves to a superposition state between sites 8 and 9, while the other half evolves to site 1 and returns back to sites 4 and 5. This is because of the relationship imposed between the mirroring times. Therefore, a MWS entanglement at $t=2t_{m,A}$ will be a given as 
\begin{equation}
    \ket{\psi(2t_{m,A})}=-\frac{1}{2}(\ket{r_4}+\ket{r_5})-\frac{i}{2}(\ket{r_8}+\ket{r_9}).
\label{maltipartite entang at 2t_m}
\end{equation}
At $t=4t_{m,A}$ the state of the system evolves to the excitation being entirely localised at site 4  
\begin{equation}
    \ket{\psi(4t_{m,A})}=\ket{r_4} \;.
\end{equation}
We note that this is not a simple return to the initial $t=0$ state. This is because there is a time-difference with respect to the excitation amplitude evolution through chains $A$ and $B$. This results in different phases arising for the amplitudes in chains $A$ and $B$, leading to destructive interference (see appendix A). Under further evolution, at $t=6t_{m,A}$, the state of the system will again form a MWS entanglement, but including a different relative phase from the one at $2t_{m,A}$ (Eq.\eqref{maltipartite entang at 2t_m}) 
\begin{equation}
    \ket{\psi(6t_{m,A})}=-\frac{1}{2}(\ket{r_4}+\ket{r_5})+\frac{i}{2}(\ket{r_8}+\ket{r_9}).
\label{maltipartite entang at 6t_m} 
\end{equation}
At $t=8t_{m,A}$ the state finally evolves back to its initial state 
\begin{equation} 
    \ket{\psi(8t_{m,A})}=\ket{r_5}.   
\end{equation}

The detailed periodicity of our SN system dynamics can be seen by plotting the evolution of the excitation probability for each site as a function of time, as shown in Fig.\eqref{fidelities of sites 1,4,5,8,9}.

\begin{figure}[ht!]
    \centering```````````
    \includegraphics[width=0.45\textwidth]{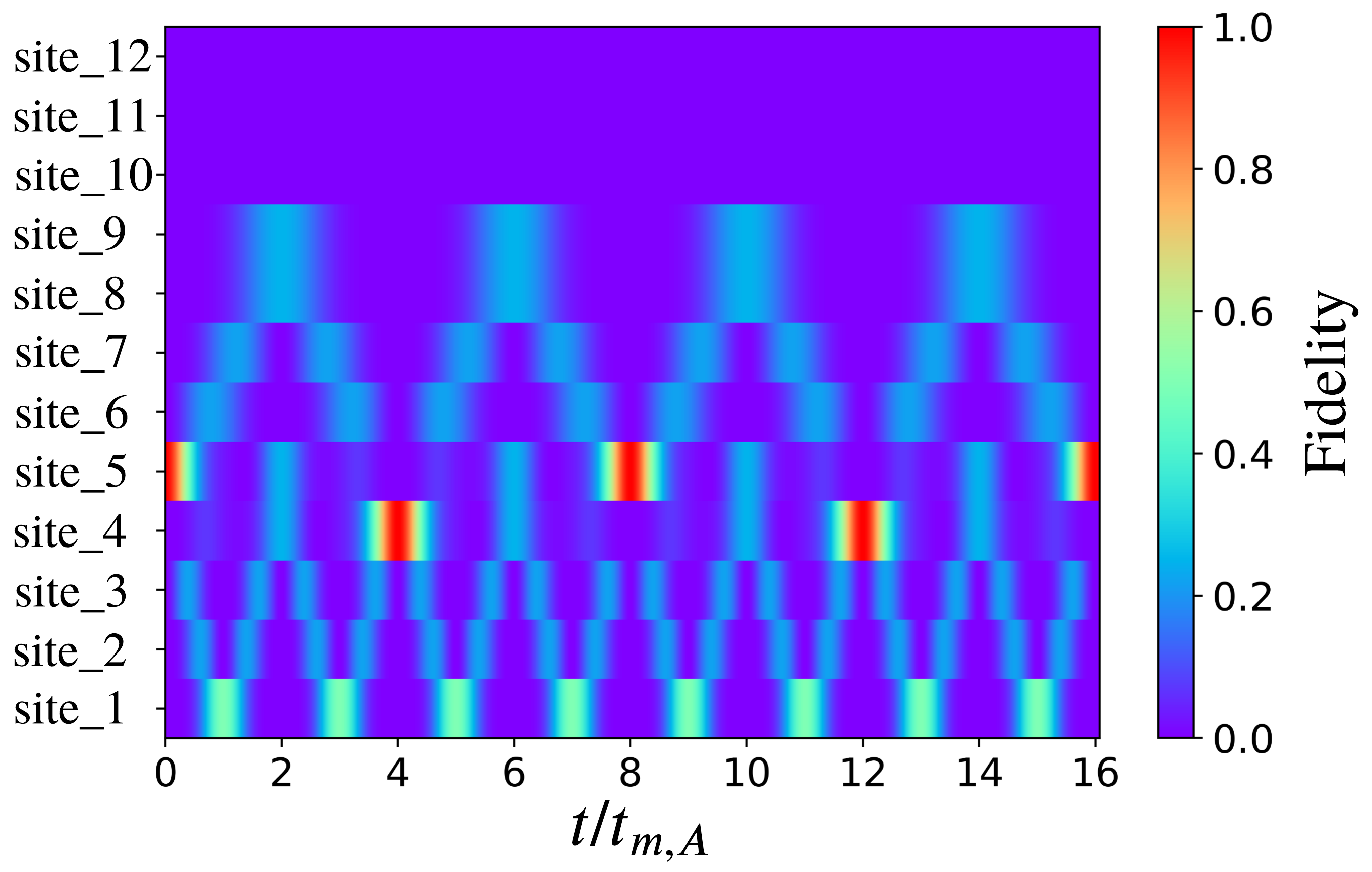}
    \caption{Fidelity of each site as a function of the rescaled time $t/t_{m,A}$. It is clear that the MWS entanglement (between sites 4, 5, 8, and 9) is periodic.} 
    \label{fidelities of sites 1,4,5,8,9}.     
\end{figure} 

\begin{figure}[H]
    \centering
    \includegraphics[width=0.5\textwidth]{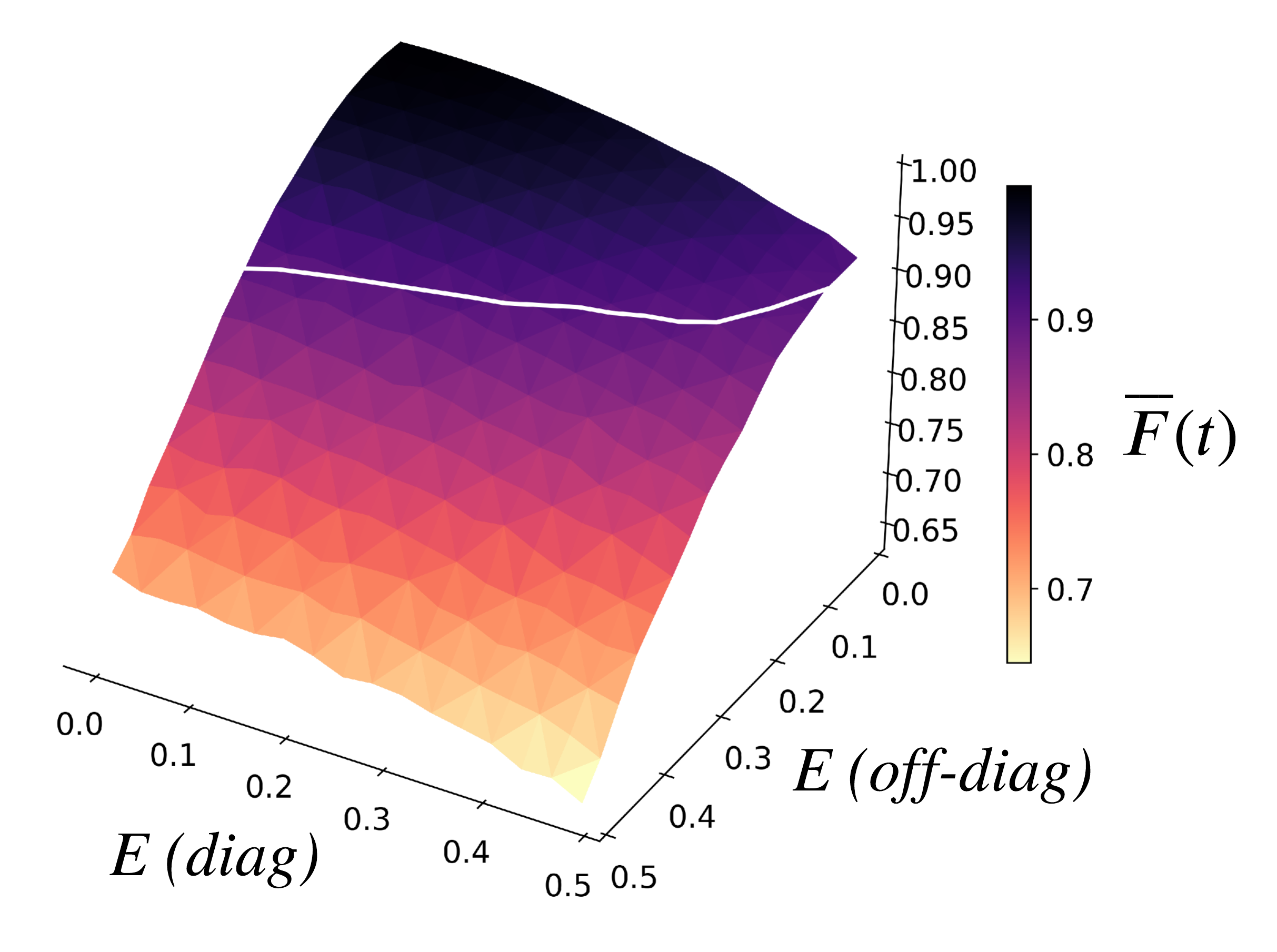}
    \caption{The robustness of the MWS entanglement in the presence of diagonal disorder (\textit{diag}) and off-diagonal disorder (\textit{off-diag}) with different error strengths, $E$, ranging from 0 to $50\%$. Each point has been averaged over 1000 realisations. White line has same meaning as in Fig.~\ref{router robustness vs N and vs disorders}.
    }%
    \label{ME state robustness}.     
\end{figure}

Let us now investigate the robustness of the MWS entanglement generated at $t=2t_{m,A}$, Eq.\eqref{maltipartite entang at 2t_m}, against both types of disorder. Fig.\eqref{ME state robustness} shows that the fidelity of the MWS entanglement is very robust against both types of disorders, particularly against diagonal disorder.

\subsubsection{Maximally Entangled State}
The variation of $J_{max}$ also enables generation of a bipartite maximally entangled state between both ends of the 3-chain SN, Fig.\eqref{fig:20}, at a specific time. This is not possible in the SN of three {\it identical} spin chains (i.e., SN of equal $J_{max}$) because of the timings of the amplitude propagation. Therefore, the maximum coupling of the second chain is set to be $J_{max,B}=1/2$ and as a result, $t_{m,B}=2t_{m,A}$. 

We start by injecting a single-excitation at site 5 at $t=0$, $\ket{\psi(0)}=\ket{r_5}$ and injecting a phase flip $e^{i\pi}$ at site 9 at $t=2t_{m,A}$, with the result that the further evolved state of the system at $t=3t_{m,A}$ will be given by 
\begin{equation}
    \ket{\psi(3t_{m,A})}=\frac{-1}{\sqrt{2}}(i\ket{r_1}-\ket{r_{12}})\;.
\end{equation} 

\begin{figure*}[ht!]
    \includegraphics[width=0.8\textwidth]{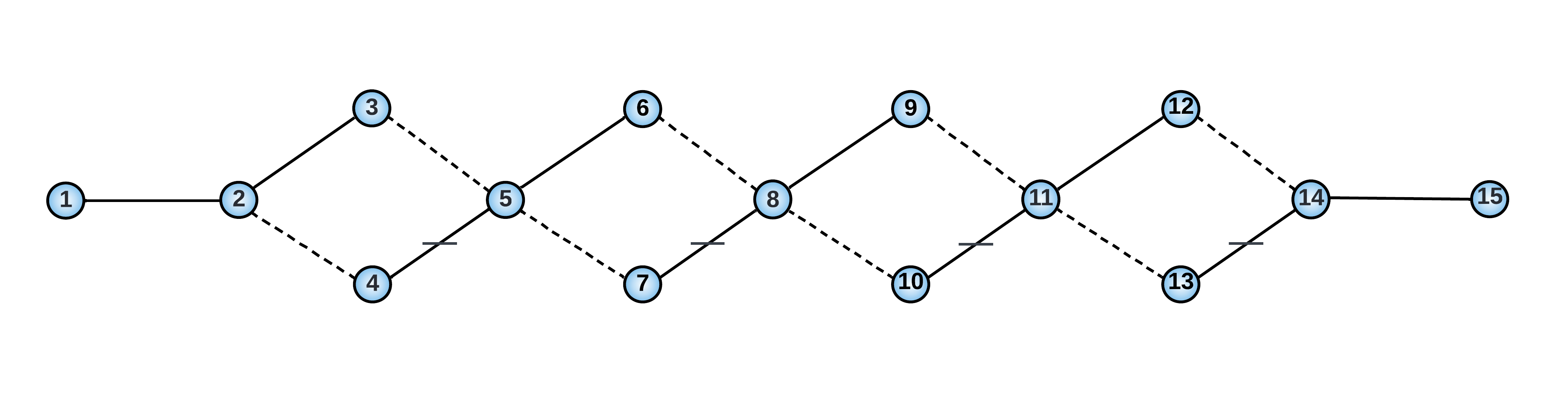}
    \caption{Diagram of a large SN.} 
    \label{fig:5ch_SN_3sites}.     
\end{figure*} 

This shows how variation of $J_{max}$ values in parts of a complex SN can adjust the timing of amplitude arrivals, to deliver all parts of a desirable distributed state at the same time.

\subsection{Example: $M$ chains of 3-sites SN}
Here, we illustrate the possibility of generalising the results to even more complex and larger SNs. The building blocks of the SN discussed here are uncoupled 3-site PST chains,  coupled together with a unitary transformation, \cite{alsulami2022unitary}, as shown in Fig.\eqref{fig:5ch_SN_3sites} for the case of $M=5$. 

Operating this SN as a router by sending an excitation from site 1 to site 15 is possible using an extension of the router protocol discussed above. It can be achieved with a single-excitation injected at site 1 at $t=0$ and a phase flip applied separately at site 4 at $t_m$, at site 7 at $2t_m$, at site 10 at $3t_m$, and at site 13 at $4t_m$.

Generation of a W-state entanglement is also possible  following the protocol shown in Fig.\eqref{W entanglement protocol in the 9s SN}. 

\begin{figure*}[ht!]
    \centering
    \includegraphics[width=0.8\textwidth]{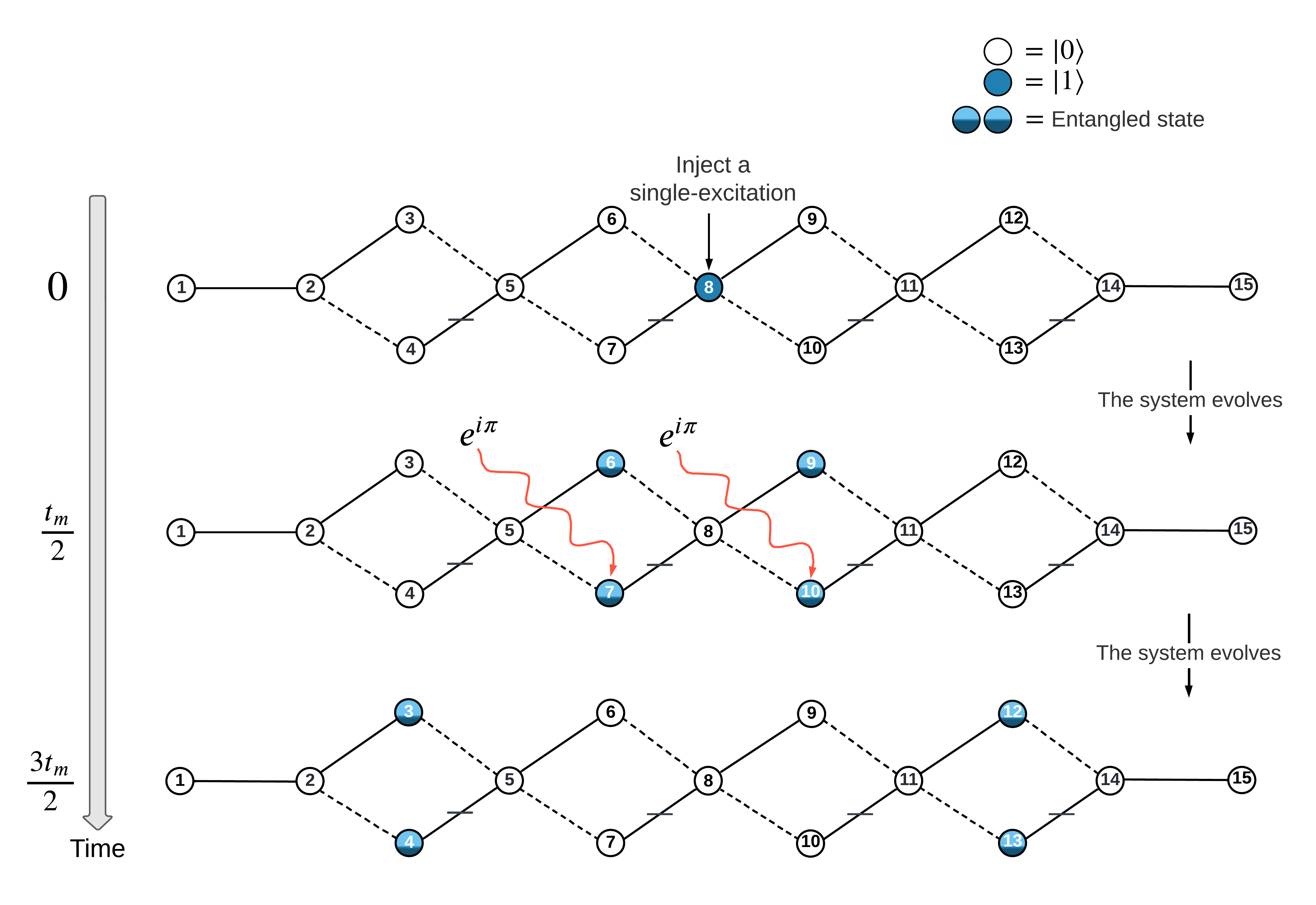}
    \caption{Generation of MWS entanglement between further sites. At $t=0$, we inject a single-excitation at site 8 and evolve for $t_m/2$, then we apply two phases flip simultaneously at sites 7 and 10, then we evolve for $t_m$ where the state will be a MWS entanglement between sites 3, 4, 12, and 13.} 
    \label{multipartiteprotocol_4chains_SN2}.     
\end{figure*}

\subsubsection{Generation of MWS entanglement}
MWS entanglement can be generated between any four sites at the vertices of the diamonds of the SN in Fig.\eqref{fig:5ch_SN_3sites}, due to the richer topology of this SN. 
The MWS entanglement can be generated, between sites 3, 4, 6, and 7 if we inject a single-excitation at site 5 and evolve for $t_m/2$, whereas if we inject a single-excitation at site 8 and evolve for $t_m/2$, then the MWS entanglement will be generated between sites 6, 7, 9, and 10. Similarly, the MWS entanglement can be generated if the single-excitation is injected at site 11. It is also possible to generate the MWS entanglement between sites 3, 4, 9, and 10 if a single-excitation is injected at either sites 6 or 7 and evolved for $t_m$. Similarly, if we inject a single-excitation at either sites 9 or 10 and evolve for $t_m$, a MWS entanglement between sites 6, 7, 12, and 13 will be generated. This method of generating MWS entanglement is applicable to longer 3-site-chains' SN systems in the same way. 

\subsubsection{MWS entanglement transfer}
\begin{figure}[ht!]
    \includegraphics[width=0.45\textwidth]{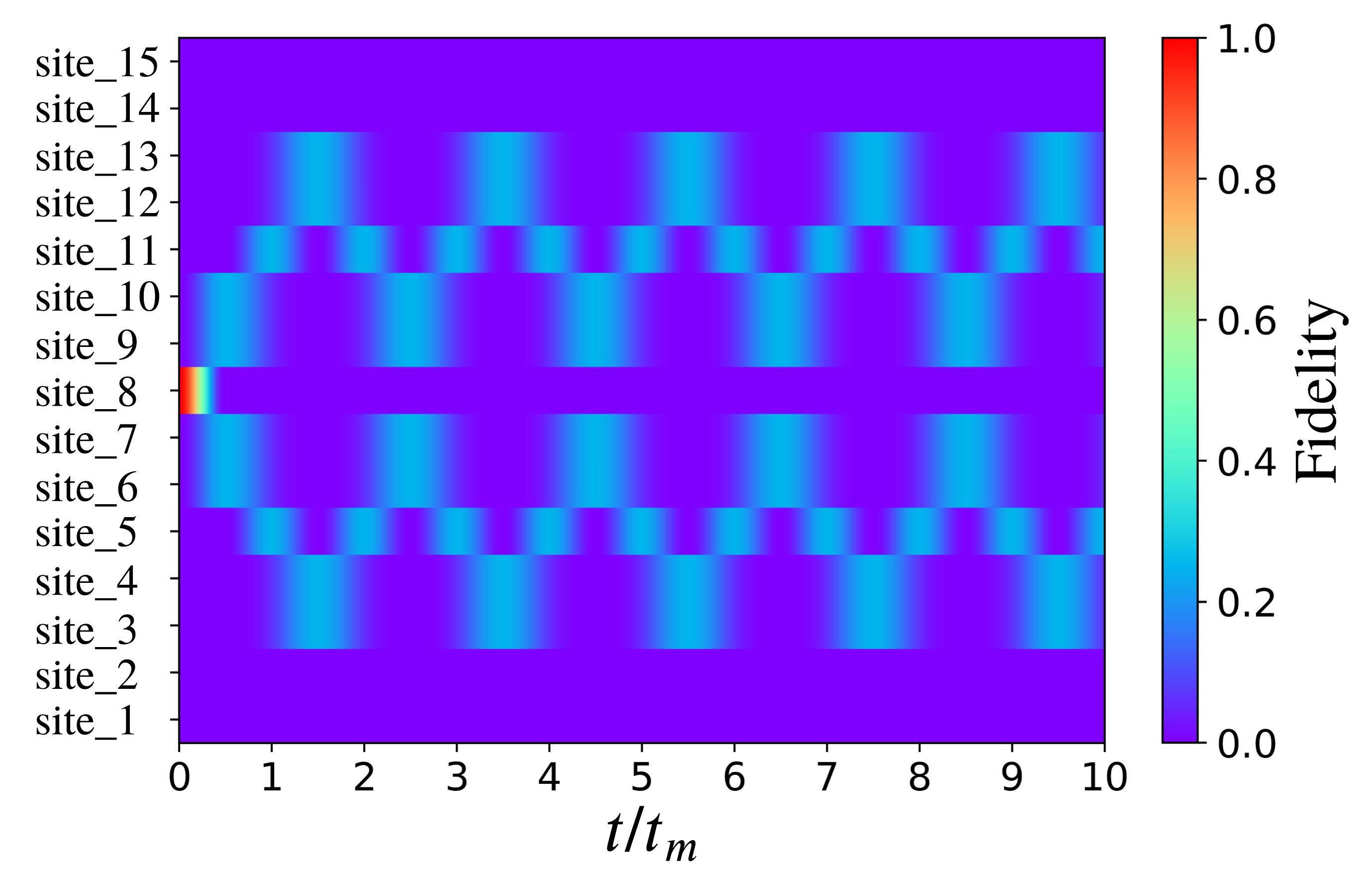}
    \caption{The fidelity of each site as a function of the rescaled time $t/t_m$ for the protocol shown in Figure.\eqref{multipartiteprotocol_4chains_SN2}. It is clear that the state keeps evolving from a MWS entanglement (between sites 6, 7, 9, and 10) to another MWS entanglement (between sites 3, 4, 12, and 13), and vice versa.} 
    \label{fig:allsites5chains_SN}.     
\end{figure}

This SN allows also to  transfer the MWS entanglement through the SN by applying a phase to generate quantum interference and change the spatial direction of the excitation evolution. The related protocol  is illustrated in Fig.\eqref{multipartiteprotocol_4chains_SN2}. Due to the PST in the building blocks of our SN, the excitation will keep oscillating between MWS entanglement of sites 3, 4, 12, and 13 and MWS entanglement of sites 6, 7, 9, and 10. This is illustrated in Fig.\eqref{fig:allsites5chains_SN} with plotting the fidelity of each site as a function of time. 

\begin{figure}[ht!]
    \centering
    \includegraphics[width=0.45\textwidth]{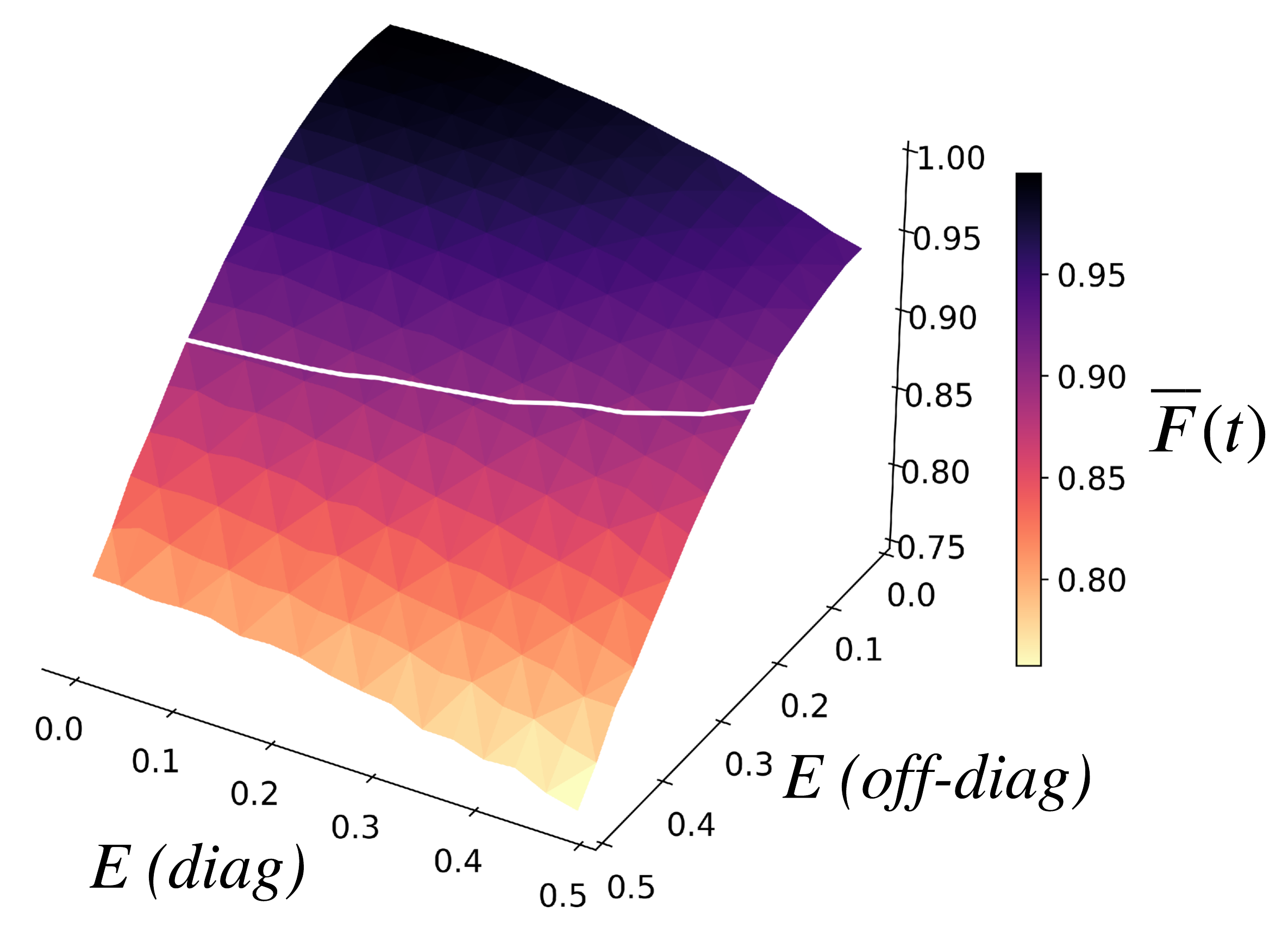}
    \caption{The robustness of the fidelity of the MWS entanglement at $3t_m/2$, in the presence of diagonal disorder (\textit{diag}) and off-diagonal disorder (\textit{off-diag}) with different error strengths, $E$, ranging from 0 to $50\%$. Each point has been averaged over 1000 realisations. White line has same meaning as in Fig.~\ref{router robustness vs N and vs disorders}.
    }%
    \label{MultiPartite robustness vs err}.     
\end{figure}

The robustness of the MWS entanglement is investigated by measuring the fidelity of the system against a desirable state chosen as a MWS entanglement . The fidelity is measured at $3t_m/2$ (i.e., at the first time the MWS entanglement is formed between sites 3, 4, 12, and 13) in the presence of two types of disorder, Fig\eqref{MultiPartite robustness vs err}.  
The MWS entanglement robustness against diagonal disorder is shown to have fidelity $>99\%$ up to large error strength of $E=10\%$. The fidelity is very robust up to error strength of $E=5\%$ for both types of disorder. Since in real experiments the error strength is expected to be much less than $10\%$, our SN can be used to generate a MWS entanglement with very good robustness against disorder. The robustness and scalability in generating such entangled states give our SN the potential to be used in quantum teleportation in solid-state systems.

\clearpage
\section{Conclusion}
\label{Conclusion}
Utilising unitary transformations to couple SN components \cite{alsulami2022unitary}, we have designed and modelled a range of scalable SN systems, capable of various quantum information processing tasks. We have realised and investigated in detail systems capable of routing and unknown phase sensing. In addition, we have shown how components of different size can be coupled together, to achieve coordinated behaviour at a chosen later time, by control of the maximum coupling between qubits, $J_{max}$, in the various components.

We have also designed and demonstrated SN systems capable of creating and distributing bipartite, W-state and multi-party W-state entanglement. We have shown that the W-state entanglement generation protocol can be used in any SN comprising an even or odd number of chains, as long as the numbers of sites in each relevant chain are equal. Otherwise, $J_{max}$ adjustments can be applied, to tune mirror times $t_m$, in order to achieve the desired distributed final state at a chosen later time.

All our SN investigations have been undertaken including the presence of diagonal and off-diagonal disorder. Modelling of these large SN systems has demonstrated significant robustness, even with large error strength $E$. In general, SN robustness is observed up to a typical error strength of $E\leq10\%$. For example, the fidelity of the router protocol in SN systems remains above $92\%$ with $E=10\%$, for chain sizes up to $N=100$ subject to diagonal disorder. It also remains above $90\%$ with $E=10\%$ up to $N=40$, subject to the more damaging off-diagonal disorder. All our robustness results are very encouraging for future experimental applications, because $10\%$ is a relatively high error strength.  Good realistic devices can be expected to have error strengths significantly lower than $10\%$. We also note that both methods of scaling the SN (increasing the number of sites per chain whilst keeping the number of chains fixed, or vice versa) show very similar robustness against disorder. (We do not present detailed numerical results here.) 




Our work presented thus far has been undertaken considering just the single-excitation sub-space of the SN systems. However, considering two- or multiple-excitation sub-spaces has significant potential to expand the range of possible applications of such SN systems, for example, the engineering of unitary design for quantum gates and protocols, leveraging the  higher excitation sub-spaces. We plan to explore such applications in future work.

\label{Conclusion}

\section{Acknowledgment} 
AHA acknowledges support from the WW Smith Fund at the University of York.

\appendix
\section{Excitation amplitudes evolving different phases}
\label{appendix A}
When a single-excitation is injected at site 5 (Fig.\eqref{fig:20} in Section \ref{Sec. MWS entang}), amplitudes for the excitation will evolve through both chains of the SN. Thus, an amplitude for the excitation evolves through the chain A of length $N_A$ and another amplitude evolves through the chain B of length $N_B$. The different phases that result from the excitation amplitudes evolving though A and B determine the final state, as we now describe.           

Consider the excitation amplitude that evolves through A. This will evolve to site 1 at $t_{m,A}$ with an overall phase factor of $e^{-i\alpha(N_A)}=(-i)^{N_A-1}=i$, as $N_A=4$. Thus, the excitation amplitude at site 1 at $t_{m,A}$ will be given by
\begin{equation}
    \ket{site_1(t_{m,A})}=\frac{i}{\sqrt{2}}\ket{r_1}.
\end{equation}

Then, this excitation amplitude evolves back from site 1 to being in a superposition state between sites 4 and 5, acquiring another overall phase of $i$, 
so the amplitude at $2t_{m,A}$ will now be given with an overall phase of -1, 

\begin{equation}
\label{N_A, t_m}
    \ket{site_{4,5}(2t_{m,A})}=-\frac{1}{2}(\ket{r_4}+\ket{r_5}). 
\end{equation}
If the system evolves for another period of $2t_{m,A}$, the amplitude at $4t_{m,A}$ will be given by 
\begin{equation}
\label{N_A state}
    \ket{site_{4,5}(4t_{m,A})}=\frac{1}{2}(\ket{r_4}+\ket{r_5}). 
\end{equation}

Consider now the other amplitude of the excitation, that evolves through chain B, which by design has weaker couplings. This will therefore take twice the time the excitation amplitudes takes to evolve through chain A, so $t_{m,B}=2t_{m,A}$. 
The amplitude of the single excitation injected at site 5 that evolves through the chain B ends up in a superposition state between sites 8 and 9 at $t_{m,B}$, with an overall phase factor of $(-i)^{N_B-1}=i$, but in addition because the coupling between sites 5 and 6 is negative, the overall phase is $-i$, 
\begin{equation}
    \ket{site_{8,9}(t_{m,B})}=-i\frac{1}{2}(\ket{r_8}+\ket{r_9}). 
\end{equation}
Following this, the excitation amplitude will evolve back to being in a superposition state between sites 4 and 5 at $2t_{m,B}$, with another overall phase of $i$ and a further relative phase of $-1$ at site 5 because of the negative coupling between sites 5 and 6. Thus, the amplitude at $2t_{m,B}$ will be given by 
\begin{equation}
\label{N_B state}
    \ket{site_{4,5}(2t_{m,B})}=\frac{1}{2}(\ket{r_4}-\ket{r_5}). 
\end{equation}

We have seen how the excitation amplitudes evolve through each chain. These can be combined to give the state of the system at $2t_{m,B}$, which will simply be the sum of equations Eq.\eqref{N_A state} and Eq.\eqref{N_B state}, as $2t_{m,B}=4t_{m,A}$, so the state of the system at $2t_{m,B}$ is given as 
\begin{equation}
\label{final state}
    \ket{\psi(2t_{m,B}}=\frac{1}{2}(\ket{r_4}+\ket{r_5}+\ket{r_4}-\ket{r_5})=\ket{r_4}. 
\end{equation}
It is now clear that if there is no difference in the excitation time through each chain (i.e., $t_{m,A}=t_{m,B}$), then the state of the system at $2t_{m,B}$ will simply be the sum of Eq.\eqref{N_A, t_m} and Eq.\eqref{N_B state}, as $2t_{m,B}=2t_{m,A}$, so the state would be given as 
\begin{equation}
\label{final2 state}
    \ket{\psi(2t_{m,B}}=\frac{1}{2}(-\ket{r_4}-\ket{r_5}+\ket{r_4}-\ket{r_5})=-\ket{r_5}. 
\end{equation}
We note that if we evolve for another $2t_{m,B}$, then both equations (Eq.\eqref{final state} and Eq.\eqref{final2 state}) will evolve back to $\ket{r_5}$, demonstrating periodicity with period $4t_{m,B}$. 

\clearpage

\bibliographystyle{unsrt}   
\bibliography{reff} 

\begin{thebibliography}{10}

\bibitem{alsulami2022unitary}
Abdulsalam~H Alsulami, Irene D'Amico, Marta~P Estarellas, and Timothy~P Spiller.
\newblock {\em Advanced Quantum Technologies}, 5(8):2200013, 2022.

\bibitem{bose2007quantum}
Sougato Bose.
\newblock {\em Cont. Phys.}, 48(1):13--30, 2007.

\bibitem{nikolopoulos2004electron}
Georgios~M Nikolopoulos, David Petrosyan, and P~Lambropoulos.
\newblock {\em Journal of Physics: Condensed Matter}, 16(28):4991, 2004.

\bibitem{serra2021pretty}
Pablo Serra, Alejandro Ferr{\'o}n, and Omar Osenda.
\newblock {\em arXiv preprint arXiv:2101.03194}, 2021.

\bibitem{christandl2004perfect}
Matthias Christandl, Nilanjana Datta, Artur Ekert, and Andrew~J Landahl.
\newblock {\em Phys. Rev. Lett.}, 92(18):187902, 2004.

\bibitem{burgarth2005conclusive}
Daniel Burgarth and Sougato Bose.
\newblock {\em Physical Review A}, 71(5):052315, 2005.

\bibitem{vinet2012construct}
Luc Vinet and Alexei Zhedanov.
\newblock {\em Physical Review A}, 85(1):012323, 2012.

\bibitem{ramirez2015entanglement}
Giovanni Ram{\'\i}rez, Javier Rodr{\'\i}guez-Laguna, and Germ{\'a}n Sierra.
\newblock Entanglement over the rainbow.
\newblock {\em Journal of Statistical Mechanics: Theory and Experiment}, 2015(6):P06002, 2015.

\bibitem{d2007freezing}
Irene D’Amico, Brendon~W Lovett, and Timothy~P Spiller.
\newblock {\em Phys. Rev. A.}, 76(3):030302, 2007.

\bibitem{estarellas2017robust}
Marta~P Estarellas, Irene D'Amico, and Timothy~P Spiller.
\newblock {\em Phys. Rev. A.}, 95(4):042335, 2017.

\bibitem{loss1998quantum}
Daniel Loss and David~P DiVincenzo.
\newblock {\em Phys. Rev. A.}, 57(1):120, 1998.

\bibitem{d2006quantum}
Irene D’Amico.
\newblock {\em Microelectron. J.}, 37(12):1440--1441, 2006.

\bibitem{brown2016co}
Kenneth~R Brown, Jungsang Kim, and Christopher Monroe.
\newblock {\em In: npj Quantum Information}, 2(1):16034, 2016.

\bibitem{berggren2004quantum}
Karl~K Berggren.
\newblock {\em Proceedings of the IEEE}, 92(10):1630--1638, 2004.

\bibitem{karamlou2022quantum}
Amir~H Karamlou, Jochen Braum{\"u}ller, Yariv Yanay, Agustin Di~Paolo, Patrick~M Harrington, Bharath Kannan, David Kim, Morten Kjaergaard, Alexander Melville, Sarah Muschinske, et~al.
\newblock {\em npj Quantum Information}, 8(1):1--8, 2022.

\bibitem{mortimer2021evolutionary}
Luke Mortimer, Marta~P Estarellas, Timothy~P Spiller, and Irene D'Amico.
\newblock {\em Advanced Quantum Technologies}, 4(8):2100013, 2021.

\bibitem{ChristandlMatthias2005Ptoa}
Matthias Christandl, Nilanjana Datta, Tony~C Dorlas, Artur Ekert, Alastair Kay, and Andrew~J Landahl.
\newblock {\em Phys. Rev. A.}, 71(3), 2005.

\bibitem{KayAlastair2011Bopc}
Alastair Kay.
\newblock {\em Phys. Rev. A.}, 84(2), 2011.

\bibitem{pemberton2011perfect}
Peter~J Pemberton-Ross and Alastair Kay.
\newblock {\em Physical Review Letters}, 106(2):020503, 2011.

\bibitem{roy2018response}
Sudipto~Singha Roy, Himadri~Shekhar Dhar, Debraj Rakshit, Aditi Sen, Ujjwal Sen, et~al.
\newblock {\em Physical Review A}, 97(5):052325, 2018.

\bibitem{riegelmeyer2021generation}
Jan Riegelmeyer, Dan Wignall, Marta~P Estarellas, Irene D’Amico, and Timothy~P Spiller.
\newblock {\em Quantum Information Processing}, 20(1):1--20, 2021.

\bibitem{degen2017quantum}
Christian~L Degen, F~Reinhard, and Paola Cappellaro.
\newblock {\em Rev. of Mod. Phys.}, 89(3):035002, 2017.

\bibitem{pirandola2018advances}
Stefano Pirandola, B~Roy Bardhan, Tobias Gehring, Christian Weedbrook, and Seth Lloyd.
\newblock {\em Nature Photonics}, 12(12):724--733, 2018.

\bibitem{wojcik2005unmodulated}
Antoni Wojcik, Tomasz Luczak, Pawel Kurzynski, Andrzej Grudka, Tomasz Gdala, and Malgorzata Bednarska.
\newblock Unmodulated spin chains as universal quantum wires.
\newblock {\em Physical Review A}, 72(3):034303, 2005.

\bibitem{oh2012effect}
Sangchul Oh, Yun-Pil Shim, Jianjia Fei, Mark Friesen, and Xuedong Hu.
\newblock {\em Phys. Rev. B.}, 85(22):224418, 2012.

\bibitem{banchi2010optimal}
Leonardo Banchi, Tony John~George Apollaro, Alessandro Cuccoli, Ruggero Vaia, and Paola Verrucchi.
\newblock {\em Phys. Rev. A.}, 82(5):052321, 2010.

\bibitem{banchi2011nonperturbative}
Leonardo Banchi, Abolfazl Bayat, Paola Verrucchi, and Sougato Bose.
\newblock {\em Phys. Rev. Lett.}, 106(14):140501, 2011.

\bibitem{kay2010perfect}
Alastair Kay.
\newblock {\em International Journal of Quantum Information}, 8(04):641--676, 2010.

\bibitem{karbach2005spin}
Peter Karbach and Joachim Stolze.
\newblock {\em Phys. Rev. A.}, 72(3):030301, 2005.

\bibitem{kostak2007perfect}
V~Kostak, GM~Nikolopoulos, and I~Jex.
\newblock {\em Phys. Rev. A.}, 75(4):042319, 2007.

\bibitem{wootters2001entanglement}
William~K Wootters.
\newblock {\em Quantum Inf. Comput.}, 1(1):27--44, 2001.

\bibitem{ronke2016anderson}
Rebecca Ronke, Marta~P Estarellas, Irene D’Amico, Timothy~P Spiller, and Takayuki Miyadera.
\newblock {\em The European physical journal. D, Atomic, molecular, and optical physics}, 70(9), 2016.

\bibitem{moehring2007entanglement}
David~L Moehring, Peter Maunz, Steve Olmschenk, Kelly~C Younge, Dzmitry~N Matsukevich, L-M Duan, and Christopher Monroe.
\newblock {\em Nature}, 449(7158):68--71, 2007.

\bibitem{stephenson2020high}
LJ~Stephenson, DP~Nadlinger, BC~Nichol, S~An, P~Drmota, TG~Ballance, K~Thirumalai, JF~Goodwin, DM~Lucas, and CJ~Ballance.
\newblock {\em Physical review letters}, 124(11):110501, 2020.

\bibitem{krutyanskiy2023entanglement}
V~Krutyanskiy, M~Galli, V~Krcmarsky, S~Baier, DA~Fioretto, Y~Pu, A~Mazloom, P~Sekatski, M~Canteri, M~Teller, et~al.
\newblock {\em Physical Review Letters}, 130(5):050803, 2023.

\bibitem{almeida2016quantum}
Guilherme~MA Almeida, Francesco Ciccarello, Tony~JG Apollaro, and Andre~MC Souza.
\newblock Quantum-state transfer in staggered coupled-cavity arrays.
\newblock {\em Physical Review A}, 93(3):032310, 2016.

\bibitem{gualdi2011modular}
Giulia Gualdi, Salvatore~M Giampaolo, and Fabrizio Illuminati.
\newblock Modular entanglement.
\newblock {\em Physical Review Letters}, 106(5):050501, 2011.

\bibitem{guhne2009entanglement}
Otfried G{\"u}hne and G{\'e}za T{\'o}th.
\newblock {\em Physics Reports}, 474(1-6):1--75, 2009.

\bibitem{m2019tripartite}
M{\'a}rcio M.~Cunha, Alejandro Fonseca, and Edilberto O.~Silva.
\newblock {\em Universe}, 5(10):209, 2019.

\bibitem{ZhangChao2016Etog}
Chao Zhang, Cheng-Jie Zhang, Yun-Feng Huang, Zhi-Bo Hou, Bi-Heng Liu, Chuan-Feng Li, and Guang-Can Guo.
\newblock {\em Scientific reports}, 6(1):39327--39327, 2016.

\bibitem{shi2002teleportation}
Bao-Sen Shi and Akihisa Tomita.
\newblock {\em Physics Letters A}, 296(4-5):161--164, 2002.

\bibitem{zhou2018efficientdense}
You-Sheng Zhou, Feng Wang, and Ming-Xing Luo.
\newblock {\em International Journal of Theoretical Physics}, 57(7):1935--1941, 2018.

\bibitem{chen2008controlled}
Xiu-Bo Chen, Qiao-Yan Wen, Fen-Zhuo Guo, Ying Sun, Gang Xu, and Fu-Chen Zhu.
\newblock {\em International Journal of Quantum Information}, 6(04):899--906, 2008.

\bibitem{walter2016multipartite}
Michael Walter, David Gross, and Jens Eisert.
\newblock {\em Quantum Information: From Foundations to Quantum Technology Applications}, pages 293--330, 2016.

\end{thebibliography}

\end{document}